\documentclass[a4paper,11pt]{article}
\pdfoutput=1 %

\usepackage{jheppub} %

\usepackage[T1]{fontenc} %

\usepackage{amsmath}
\usepackage{amssymb}
\usepackage{amsthm}
\usepackage{mathrsfs}
\usepackage{physics}
\usepackage{bbm}
\usepackage{enumitem}
\usepackage{cancel}

\newtheorem{theorem}{Theorem}
\newtheorem{lemma}{Lemma}
\newtheorem*{lemma*}{Lemma}

\theoremstyle{definition}

\theoremstyle{plain}

\theoremstyle{remark}
\newtheorem{remark}{Remark}
\usepackage{tikz}
\usepackage{tikz-cd}
\usepackage{tikz-3dplot}

\title{\boldmath Proof of the R\'{e}nyi QNEC} 

\author{Thomas Faulkner}

\affiliation{Department of Physics, University of Illinois Urbana-Champaign}

\emailAdd{tomf@illinois.edu}
\abstract{
We give a general proof of the R\'{e}nyi QNEC formulated using the Araki-Masuda weighted $L_p$ norms for all $p > 2$.
This generalizes the algebraic proof of the QNEC applicable at $p=2$. Our approach utilizes a weighted version of a
conditional $L_p$ norm which extrapolates the quantum expansion/ant-quantity. We elucidate some properties
of the corresponding weighted conditional entropy both for half-sided modular inclusions and for general von Neumann subalgebras with a common cyclic and separating vector. The $p=2$ limit computes a conditional dynamical entropy originally defined in the context of black hole dynamics. 
}

\begin{document}

\maketitle
\flushbottom

\section{Introduction}
\label{sec:intro}

Consider a one-parameter family of von Neumann algebras $\mathcal{A}_V$ acting on $\mathscr{H}$ 
forming a half-sided modular inclusion (HSMI) \cite{borchers1992cpt,wiesbrock1993half,araki2005extension} based on the translation/ANEC operator $P$ \cite{Faulkner:2016mzt,Casini:2017roe} and a corresponding vacuum state $\eta$.\footnote{We use the same notation and conventions as in \cite{Faulkner:QNECwithoutAssumptions}.} This is the usual setup for
studying QFT on a curved spacetime in the presence of a Killing horizon \cite{wall2012proof} where $\mathcal{A}_V$ describes the fields restricted to the exterior of
the horizon with a cut determined by the continuous parameter $V$ and where $\eta$ is the Hartle-Hawking state. 
For some fixed pure state $\psi \in \mathscr{H}$ we would like to compute and constrain the
``sandwiched'' or ``weighted'' R\'{e}nyi relative entropy $D_p$ for $\psi$ restricted to the family $\mathcal{A}_V$. The limit $p \rightarrow 2$ gives the standard relative entropy \cite{araki1976relative,araki1977relative}, whose negative reproduces the generalized entropy along the event horizon, up to an additive constant, in the leading semiclassical gravitational approximation \cite{wall2012proof}. The generalized entropy
is the sum of the horizon area at the cut $V$, divided by $4G_N$, and the von Neumann entropy of quantum fields outside this cut \cite{Bekenstein:1973ur,Bekenstein:1974GSL,Bousso:2015mna}, a combination that has played a central role in recent developments in quantum gravity and holography \cite{Faulkner:2013ana,Jafferis:2015del,Almheiri:2014lwa,Penington:2019kki,Almheiri:2019qdq,Almheiri:2019hni,Witten:2021crossed} based on insights from quantum information theory.

Von Neumann entropies and their conditional variants are central to quantum information inferences made with access to many independent and identically prepared (iid) copies of the system of interest.  R\'{e}nyi versions of these entropies often arise when moving beyond such an iid setting. In gravitational systems generalized versions of such entropies (mostly focusing on the min and max conditional entropies at orders $p=\infty,1$, respectively \cite{Renner:2005qbj,Koenig:2009avh}) have played an important role in recent formulations of holography \cite{Akers:2020pmf,Akers:2023oneshot,Bousso:2024maxfocusing,Wang:2022islands}.
With this motivation it is natural to study and constrain the R\'{e}nyi relative entropies as possible candidates for a R\'{e}nyi version of the generalized entropy. For example, the standard relative entropy is subject to a convexity constraint along $V$, also known as the QNEC \cite{Bousso:2015mna,Bousso:2015wca,Balakrishnan:2017bjg,Koeller:2015qmn},
and in this paper we will prove that $D_p$ is convex in $V$ for general $2 \leq p \leq \infty$. Convexity is intimately tied to its identification as a generalized entropy at $p=2$.\footnote{The QNEC has also proven useful in other aspects of QFT; see, for example, \cite{Mezei:2019sla,Kibe:2021qjy,Kibe:2024icu,Casini:2023kyj}.}
The R\'{e}nyi QNEC formulated this way, in terms of R\'{e}nyi relative entropies, was originally conjectured in \cite{Lashkari:2018nsl}.

Previous work has proven the R\'{e}nyi QNEC for specific QFTs, namely free quantum field theory \cite{Moosa:2020jwt,Roy:2022RenyiQNEC}.  The paper \cite{Moosa:2020jwt} provided counterexamples for the case $1 \leq p < 2$
so in this paper we focus only on $p \geq 2$.
A recent paper also found a proof valid for any QFT \cite{Kibe:2026wsg}, using the same HSMI machinery as in this paper, but for specific values of the R\'{e}nyi index $\alpha \in \mathbb{Z}_{\geq 2}$ where the index is $\alpha = p/2$.
The authors of that paper introduced many interesting and powerful methods for this problem; however the relationship between these methods and the general algebraic proof at $p=2$ \cite{Ceyhan:2018zfg} based on the ant conjecture \cite{Wall:2017blw} is not at all transparent. One would hope for a unifying proof whose methods have some sort of ``continuity'' with the $p \rightarrow 2$ limit.
This paper will achieve this, albeit with the addition of some new machinery that was not necessary at $p=2$. 

In particular we will introduce a new weighted conditional R\'{e}nyi entropy where the weighting arises from a common cyclic and separating vector. Properties of this weighted conditional entropy constrain the non-conditional R\'{e}nyi relative entropies in certain limits. For example, the limit $p \rightarrow 2$ reproduces a generalized dynamical entropy quantity, studied in \cite{Hollands:2024dynamical}, that tracks the evolution of the generalized entropy of an apparent horizon for a perturbed static black hole. It is known that monotonicity of this entropy is related to the QNEC.

Note that the name ``R\'{e}nyi QNEC'', usually understood as a statement about the second-order shape variation of R\'{e}nyi relative entropies, here refers exclusively to its diagonal shape variations (in $d=2$ the diagonal variation is everything).
The positivity of the second-order \emph{off-diagonal} shape variations of the standard relative entropy for null cuts of a Killing
horizon was proven in \cite{Ceyhan:2018zfg,Casini:2017roe} and is generally known as strong superadditivity. Due to special properties of such null cuts, and under some regularization scheme, this result also derives from strong subadditivity of von Neumann entropy \cite{Bousso:2015mna}.
Counterexamples to the off-diagonal statement for the R\'{e}nyi relative entropies have been found in \cite{Kibe:2026offdiagonal}.

The situation to date seems to suffer from a lack of input from semiclassical gravity itself. Backreaction should be expected to be important when computing R\'{e}nyi-like quantities away from $p=2$ \cite{Faulkner:2013yia,Lewkowycz:2013nqa,Bao:2019RenyiRelative} even in perturbative semiclassical gravity, due to large gradients in $A/(4G_N)$ (away from stationary Killing horizons) that make up for the perturbatively small shifts from the backreaction effects. So the real surprise is that the diagonal R\'{e}nyi QNEC is even a thing. In particular the original motivation for interpreting the R\'{e}nyi QNEC as a constraint on a generalized R\'{e}nyi entropy is tenuous at best.
Instead we hope that a better understanding of the quantities introduced in this paper - the weighted R\'{e}nyi conditional entropies, R\'{e}nyi dynamical entropies, and generalized quantum expansions - and their relation to semiclassical gravity and one-shot holography will clarify this situation.

\begin{figure}[htbp]
\centering
\includegraphics[width=.8\textwidth]{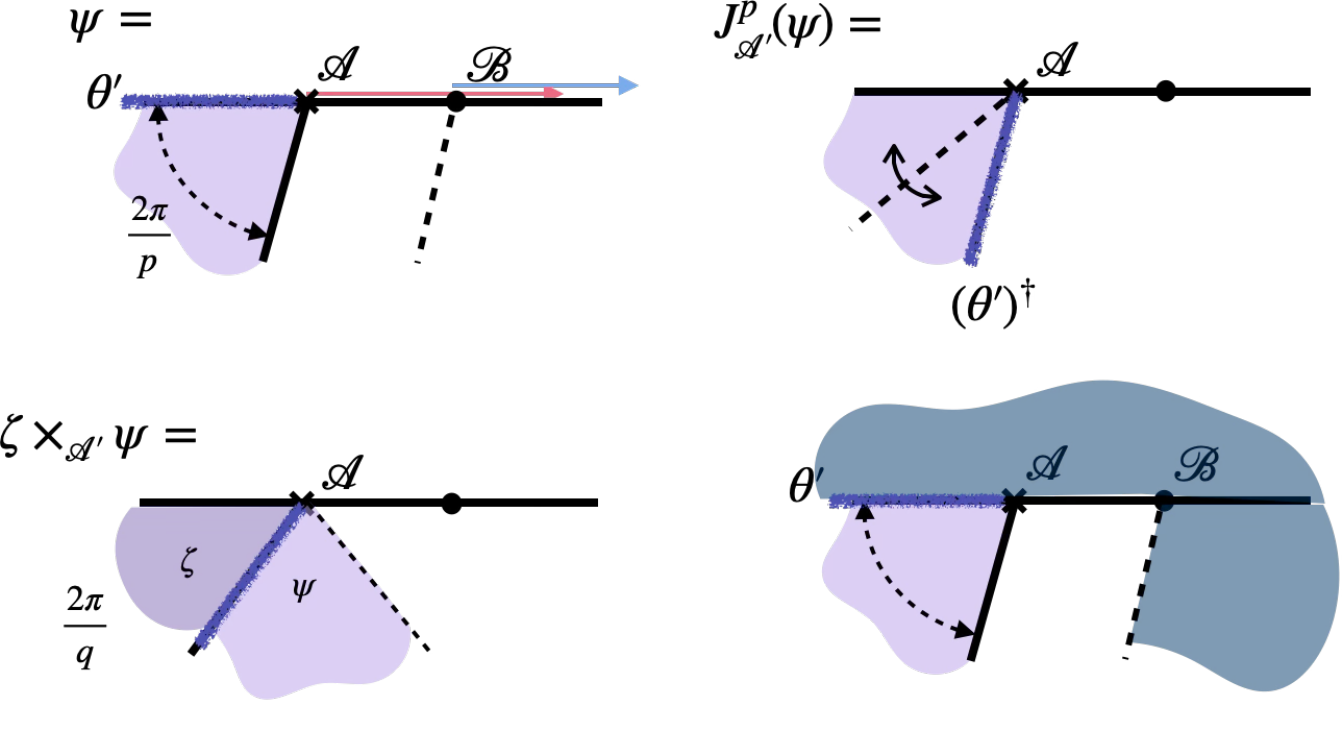}
\caption{The $L_p(\mathcal{A}'), L_p(\mathcal{B}')$ spaces can be visualized in a Euclidean plane where the real axis is the $V$ coordinate and $\mathcal{A},\mathcal{B} = \mathcal{A}_{V_1},\mathcal{A}_{V_2}$ with $V_1 < V_2$. This picture can be thought of as the Euclidean path integral
representation for $\psi$, integrating below the real axis, as a wavefunctional. No sources are allowed in the white regions. (\emph{top left}) The polar decomposition allows an isometry to act to the left at $\mathcal{A}'$. (\emph{top right}) The antilinear involution that flips across the midpoint.
(\emph{bottom left}) The H\"{o}lder product.  (\emph{bottom right}) Taking the supremum over all possible insertions in the darker region, appropriately normalized, computes the $L_p(\mathcal{B}')$ norm.  }
\label{fig:Lp}
\end{figure}

To define the R\'{e}nyi relative entropies we need the notion of a non-commutative $L_p$ norm based on the algebra $\mathcal{A}_V'$ and a cyclic and separating $\eta$ \cite{haagerup1979lp,kosaki1984applications,araki1982positive}. For a recent introduction to the Haagerup formulation aimed at QFT readers, see \cite{Kibe:2026sgf}. We will use the Araki-Masuda  \cite{araki1982positive} norms $\| \psi \|_{L_p(V')}$ which are constructed directly from the Hilbert space $\mathscr{H}$. In particular for $p \geq 2$ these are subspaces $\psi \in L_p(V') \subset \mathscr{H}$, where this norm is finite. We will give a complete definition in Section~\ref{sec:reminders}; for now see Figure~\ref{fig:Lp} for a graphical representation.
The relative entropies are  $D_p(\psi|_{V}) = \frac{2p}{p-2} \ln \| \psi \|_{L_p(V')}$ and this is the general
algebraic formulation of the sandwiched relative entropy originally defined for matrix algebras, see \cite{Berta:2016vnw,Lashkari:2018nsl,MullerLennert:2013Renyi,Wilde:2013sandwiched}. As a function of $V$, data-processing or monotonicity of relative entropy \cite{Frank:2013rov,Beigi:2013teh,Berta:2016vnw} implies that
$L_p(V_1') \subset L_p(V_2')$ for $V_1 < V_2$ and correspondingly that $D_p(\psi|_V)$ is a non-increasing function of $V$. Convexity of this function, however,
goes beyond the usual quantum information bounds.

\vspace{.5cm}

It is well known that for any $\psi \in L_p(V') $ there is a polar decomposition  $\psi = \theta'_V \Delta_{\eta,\varphi_V;V}^{-1/p} \eta$
where $ \theta'_V \in \mathcal{A}_V'$ is a partial isometry and $ \Delta_{\eta,\varphi_V;V}$ is  the relative modular operator defined using
Tomita-Takesaki relative modular theory \cite{takesaki2003theory}. Here $\varphi_V \in \mathscr{P}^\natural(V) \subset \mathscr{H}$ is in the natural positive cone
for $\mathcal{A}_V$. 
We emphasize that $\varphi_V$ plays an important role in the definition of these $L_p$ spaces.  It is not normalized; instead its norm determines the value
of the $L_p$ norm $\| \varphi_V\|^{2/p} = \| \psi \|_{ L_p(V')} $. As we will see it also determines, via correlation functions in $\varphi_V$, small perturbations of the norm. 
To further emphasize the importance of $\varphi_V$ beyond just its norm, any replica trick construction of the $L_p$ norm \cite{Lashkari:2014relative,Bao:2019RenyiRelative} would give an explicit construction
of $\varphi_V$ as an intermediate step, and in a QFT with a gravitational dual the norm could be computed after finding the gravitational dual of $\varphi_V$ \cite{Bao:2019RenyiRelative}.

An important aspect of the algebraic QNEC proof is the introduction of a quantity called the ``quantum expansion''
in \cite{Faulkner:QNECwithoutAssumptions}. As the name suggests, in certain limits of semiclassical gravity it reduces to the actual quantum corrected expansion \cite{Bousso:2015mna} up to a factor of $4 G_N$.
 This quantity captures a version of the ant conjecture in \cite{Wall:2017blw} and was first computed in \cite{Ceyhan:2018zfg}
 using Connes cocycle flow.
For various implications of this procedure in QFT and gravity see \cite{Levine:2020wedge,Bousso:2020cocycle}.
Given these considerations, the natural R\'{e}nyi analog of the ``quantum expansion'' defined in \cite{Faulkner:QNECwithoutAssumptions}  (up to a normalization that we fix in \eqref{eq:Thetap}), is given by:
\begin{equation}
\Theta_p(\psi|_V) \propto %
\inf_{ u' \in \mathcal{U}(\mathcal{A}_V')} \frac{\| \sqrt{P} u' \varphi_V \|^2}{ \| \varphi_V \|^2}
= %
\frac{1}{2} \frac{\| \sqrt{P} \varphi_V \|^2}{ \| \varphi_V \|^2}
\end{equation}
for $2 \leq p < \infty$ and we set $\Theta_\infty(\psi|_V) =0$ (see Remark~\ref{rem:pinfty}). Here $\mathcal{U}(\mathcal{A}_V')$ is the set of unitaries
contained in this von Neumann algebra, and we quote formulas in the introduction for the simplest case where all states are cyclic and separating. 
The second equality was proven in \cite{Ceyhan:2018zfg,Faulkner:QNECwithoutAssumptions} and arises since a one-sided unitary Connes cocycle can boost away half of the ANEC.

Unfortunately this quantity is not manifestly monotonic with $V$ since
$\varphi_V$ changes with $V$. At $p=2$, $\varphi_V$ is unitarily related to $\psi$ so it is manifestly monotonic from the variational form. 
Surprisingly, we will still show that $\Theta_p(\psi|_V)$ is monotonic in $V$ for all $2 \leq p \leq \infty$.
As in the regular QNEC proof we will also show that
$- \partial_V D_p(\psi|_V) = 2\pi \Theta_p(\psi|_V)$ so monotonicity of $\Theta_p$ is the statement of the R\'{e}nyi QNEC. 

As it turns out, for the R\'{e}nyi case it is convenient to introduce a one-parameter generalization of the quantum expansion
for $a > 0$ and $2 < p \leq \infty$
\begin{equation}
  \label{eq:intro-cond}
\Theta_p(\psi|_{V};a) =  - \sup_{\chi \in \mathscr{H}_{\leq 1}, u' \in \mathcal{U}(\mathcal{A}_{V}')} \frac{p}{\pi a(p-2)} \ln\frac{ \| 
e^{- %
a \sin(2\pi/p) P} u' \Delta_{\eta,\chi; V}^{1/p-1/2} \psi \|}{ \| \psi \|_{L_p(V')}}
\end{equation}
The main idea leading to the QNEC involves relating these quantities:
\begin{theorem}
\label{thm:main1}
Consider some nonzero $\psi \in L_p(V_0')$ for $ 2 < p \leq \infty$. Then, with $V \geq V_0$: 
\begin{enumerate}
\item $\Theta_p(\psi|_V ; a) \in  \mathbb{R}_{\geq 0}$ is a monotonically non-increasing function of $a$ 
with the limit:
\begin{equation}
\lim_{ a \rightarrow 0^+} \Theta_p(\psi|_V; a) = \Theta_p(\psi|_V)  
\end{equation}
\item We have the bounds:
\begin{equation}
 \Theta_p(\psi|_V) \geq \Theta_p(\psi|_V ; a) \geq \frac{D_p(\psi|_V) - D_p(\psi|_{V+a})}{2\pi a}
\end{equation}
\item The derivative exists if $V > V_0$ where it satisfies:
\begin{equation}
- \partial_V D_p(\psi|_V) = 2\pi \Theta_p(\psi|_V) < \infty
\end{equation}
\end{enumerate}
\end{theorem}

\begin{figure}[htbp]
\centering
\includegraphics[width=.8\textwidth]{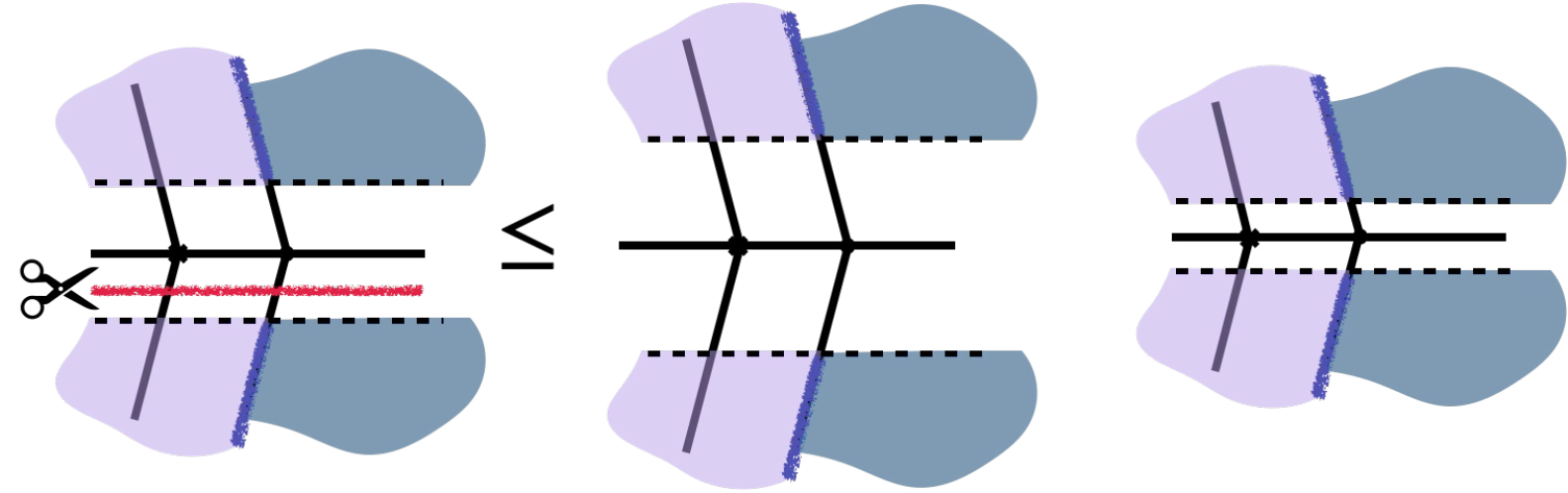}
\caption{A simple graphical argument for the monotonicity statement in Theorem~\ref{thm:main1} (1). These figures represent the new weighted conditional norm
that appears in the numerator of the argument of the log in \eqref{eq:intro-cond}. The darker regions represent the supremum over $\chi$ and the imaginary
translation $e^{- a \sin(2\pi/p) P} $ shifts these from the real axis. H\"{o}lder's inequality for the $P$ spectral integral can be given a cutting and gluing argument in the path integral much like Cauchy-Schwarz. \label{fig:holder}}
\end{figure}

\begin{figure}[htbp]
\centering
\includegraphics[width=.6\textwidth]{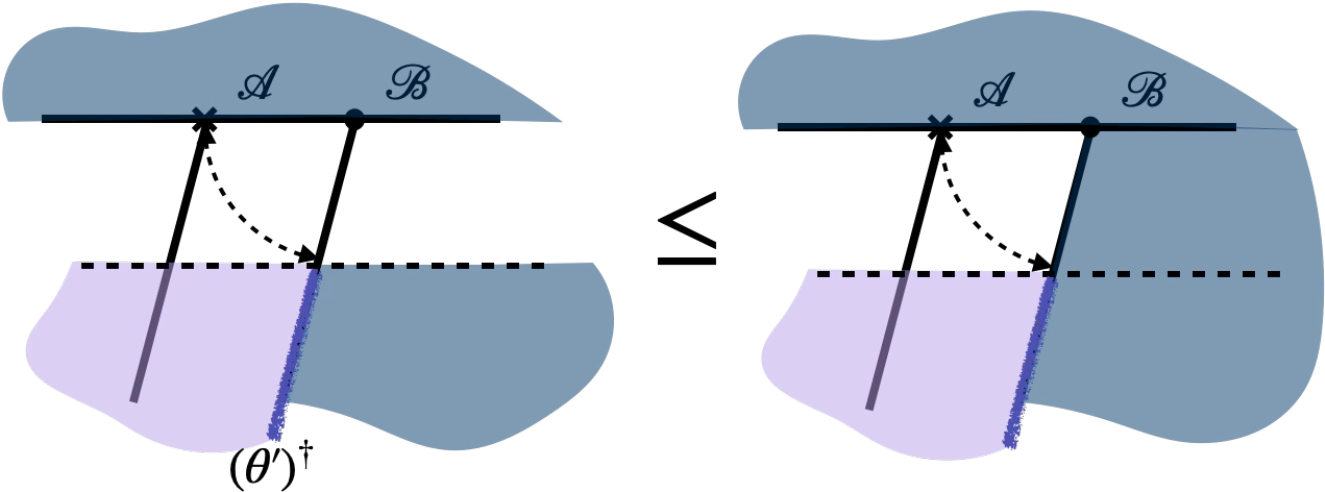}
\caption{A simple graphical argument for the second inequality in Theorem~\ref{thm:main1} (2).
 The figure on the right of the inequality can be recognized as the $L_p(\mathcal{B}')$ norm after flipping with $J^p_{\mathcal{B}'}$; see Figure~\ref{fig:Lp}.
\label{fig:conditional-bound}}
\end{figure}

The proof of this theorem can be found in Section~\ref{sec:proofs}. The outline of the proof is as follows. Monotonicity in (1) is manifest from H\"{o}lder's inequality applied to the spectral measure of $P$; see Figure~\ref{fig:holder}. The second inequality in (2) is illustrated in Figure~\ref{fig:conditional-bound}. 
The limit in (1) along with (3) can be established using a formula for the derivative of the $L_p$ norm as well as results from \cite{Ceyhan:2018zfg,Faulkner:QNECwithoutAssumptions} for the expectation value of $P$ in the cocycle-flowed state
of $\varphi_V$. While our paper was in preparation, the authors of \cite{Kibe:2026wsg} posted a v3 on the arXiv proving the integer-$p/2$ version of this ``ant-formula''.

Combining (2) and (3) gives what we will call ``left endpoint convexity''. We then show that $D_p(\psi|_V)$
is sufficiently regular (differentiable and locally Lipschitz) as a function of $V$ to extend this to full convexity. 
We summarize the R\'{e}nyi QNEC in the following:
\begin{theorem}
\label{thm:main2}
Fix $2 \leq p \leq \infty$. Consider $\psi \in L_p(V_0')$ with $\|\psi\|_{L_2}=1$ and, if $p=2$, demand $D(\psi|_{V_0}) < \infty$. Then
\begin{enumerate}
\item $V \rightarrow D_p(\psi|_{V}) \in [0,\infty)$
is a monotonically non-increasing, continuous, convex function for $V \geq V_0$ which is continuously differentiable for $V > V_0$. %
\item $V \rightarrow \Theta_p(\psi|_{V}) \in [0,\infty]$ is a monotonically non-increasing, continuous function for $V \geq V_0$.
\item They satisfy the integral relation for $V_0 \leq V_1 < V_2$:
\begin{equation}
\label{mainintegral}
D_p(\psi|_{V_1})= D_p(\psi|_{V_2})
+ 2\pi \int_{V_1}^{V_2} d V\, \Theta_p(\psi|_{V})
\end{equation}
\item For $2 \leq p'\leq p$, the maps $p' \rightarrow D_{p'}(\psi|_V), \Theta_{p'}(\psi|_{V})$ are both continuous when $V> V_0$ and respectively continuous, lower semicontinuous when $V=V_0$.
\end{enumerate} 
\end{theorem}
Note that various forms of this theorem were proven for $p=2$ in \cite{Faulkner:QNECwithoutAssumptions} and $p = 4,6, \ldots$ in \cite{Kibe:2026wsg}.

In addition to these results we also discuss a few related quantities. The generalized expansions are defined
in terms of a weighted conditional $L_p$ norm denoted $L_p(\mathcal{A}' \subset \mathcal{B}')$, generalizing some of the conditional norms discussed in \cite{Junge:2010mixednorm,Gao:2019subalgebras}
away from the case of subalgebras with conditional expectation. The associated entropies define R\'{e}nyi versions of the dynamical
conditional entropies discussed in \cite{Hollands:2024dynamical,Visser:2024dynamical,Jia:2025dynamical} and so they are quite natural in the black hole context
and as such should be of independent interest. 

We start with some reminders about R\'{e}nyi divergences in Section~\ref{sec:reminders}. Those readers familiar with the subject and comfortable guessing our notation
are encouraged to skip this section. In Section~\ref{sec:condnorms} we define the conditional norm
and elucidate some of its properties for both general subalgebras with a common cyclic and separating $\eta$, and for
HSMI. In Section~\ref{sec:expansions} we use the conditional norms to define the generalized quantum expansions. Finally in Section~\ref{sec:proofs} we complete
the proofs of the two main theorems given above. 

\section{Reminders on Weighted R\'{e}nyi Divergences}
\label{sec:reminders}

Our focus in this paper will be the weighted $L_p$ norms with $2 \leq p \leq \infty$. These will be based on a cyclic and separating
unit vector $\eta$, the dependence on which we will suppress in our notation throughout. There is a rich duality theory relating these to the $p'$ norms
with $1 \leq p' \leq 2$ that will only play a small role, so we postpone discussing this case to Section~\ref{sec:proofs} where some of these $L_{p'}$ norms will be used to compute derivatives of the R\'{e}nyi divergences and to prove certain regularity properties in $V$.

We work with a von Neumann algebra $\mathcal{A}$ in standard form $(\mathcal{A},\mathscr{H}, J_{\mathcal{A}}, \mathscr{P}^\natural(\mathcal{A}))$, and later with a subalgebra $\mathcal{B} \subset \mathcal{A}$.
 Some special properties arising from the HSMI will also be discussed and in this case our notation
will replace $\mathcal{A}_V \rightarrow V, \mathcal{A}_V' \rightarrow V'$ to lighten the load. We will follow \cite{araki1982positive} and will freely quote results from this paper and \cite{Berta:2016vnw}.

Start with $2 \leq p \leq \infty$ and define the $L_p$ space as the subset of $\mathscr{H}$ satisfying:
\begin{equation}
\psi \in L_p(\mathcal{A}') : \psi \in \bigcap_{\chi \in \mathscr{H}} \mathcal{D}( \Delta_{\eta,\chi; \mathcal{A}}^{1/p-1/2} ), \| \psi \|_{L_p(\mathcal{A}')} < \infty
\end{equation}
where the norm is
\begin{equation}
\label{Lpnorm}
\| \psi \|_{L_p(\mathcal{A}')} = \sup_{\chi \in \mathscr{H} : \| \chi \| \leq  1}  \| \Delta_{\eta,\chi; \mathcal{A}}^{1/p-1/2} \psi \|
\end{equation}

For $2 \leq p < \infty$ we have the polar decomposition for $\psi \in L_p(\mathcal{A}')$ 
\begin{equation}
\left|\psi\right> = \theta' \Delta_{\eta,\varphi;\mathcal{A}}^{-1/p} \left| \eta \right> \qquad \varphi \in \mathscr{P}^\natural(\mathcal{A})
\end{equation}
where $\theta'$ is a partial isometry $\theta' \in \mathcal{V}_\varphi(\mathcal{A}') \cap \mathcal{V}_\psi(\mathcal{A}')^\dagger$. 
We define the space
of partial isometries $\mathcal{V}_\psi(\mathcal{A}') = \{ v' \in \mathcal{A}' : (v')^\dagger v' = \pi_\psi(\mathcal{A}')\}$ where $\pi_\psi(\mathcal{A}')$
is the support projection for $\omega_\psi|_{\mathcal{A}'}$ (i.e. the smallest projection $\pi$ such that $\omega_\psi|_{\mathcal{A}'}(1-\pi) =0$).
We write $j_{\mathcal{A}}(x)=J_{\mathcal{A}}xJ_{\mathcal{A}}$ and $\widehat{\varphi}=\varphi/\|\varphi\|$ for nonzero $\varphi$.

\begin{lemma}
\label{lem:normoptimizer}
For nonzero $\psi$, in the definition of the norm \eqref{Lpnorm} we can restrict to variation over $\chi \in \mathscr{P}^\natural(\mathcal{A})_{\leq 1}$ and in this case  the supremum is uniquely achieved for $2 < p < \infty$:
\begin{equation}
\chi = \theta' j_{\mathcal{A}}(\theta') \varphi/ \| \varphi \| \quad \implies \quad
\| \psi \|_{L_p(\mathcal{A}')} = \| \varphi \|^{2/p}
\end{equation}
\end{lemma}
\begin{proof}
An application of Hadamard's three-line theorem:
\begin{equation}
\| \Delta_{\eta,\chi; \mathcal{A}}^{1/p-1/2} \psi \| =  \| \Delta_{\varphi,\chi; \mathcal{A}}^{-z}   \theta' \left| \varphi \right>  \|_{z = 1/2-1/p}
 \leq \| \pi_\chi(\mathcal{A}') \theta'  \varphi  \|^{2/p} \| (\theta')^\dagger  \chi \|^{1-2/p} \leq \|  \varphi  \|^{2/p}
\end{equation}
which saturates iff $\pi_\chi(\mathcal{A}')  = \pi_\psi(\mathcal{A}')$
and $ \left< \widehat{\varphi} \right| \Delta_{\widehat{\varphi}, (\theta')^\dagger \chi; \mathcal{A}}^{-2{\rm Re} z} \left| \widehat{\varphi} \right> = 1 $
for all $0\leq {\rm Re} z \leq 1/2$ which is possible iff 
$\chi =  \theta' j_{\mathcal{A}}(\theta') \widehat{\varphi}$ (for example the regular relative entropy comparing these two states vanishes, implying the
states are equal; the vector in the natural cone is then uniquely determined). 
\end{proof}

 The $L_p$ norm reduces to the Hilbert space norm when $p=2$ and for $p=\infty$ we
have $L_\infty(\mathcal{A}') \cong \mathcal{A}'$ as Banach spaces where $\psi = a' \eta \in L_\infty(\mathcal{A}') \leftrightarrow a'$ and $\| \psi \|_{L_\infty(\mathcal{A}')} = \| a' \|$. 
The polar decomposition for $p=\infty$ is simply the operator polar decomposition $a' = \theta' \sqrt{ (a')^\dagger a' }$.  

We recall that naturally $L_p(\mathcal{A}') \subset L_{p'}(\mathcal{A}')$ for $p \geq p' \geq 2$ as a dense subspace in the norm
of the latter.  This follows from monotonicity
of the norms $\| \psi \|_{L_p} \geq \| \psi \|_{L_{p'}}$ and density of $\mathcal{A}' \eta \cong L_\infty(\mathcal{A}')$ in $L_{p'}(\mathcal{A}')$. Monotonicity itself follows since $(1/2-1/p) \rightarrow \ln \| \psi \|_{L_p}$ is a convex map where it is defined,
and $  \| \psi \|_{L_p}  = \| \varphi \|^{2/p} \geq \| \Delta_{\eta,\varphi ;\mathcal{A}}^{-1/p} \eta \| = \| \psi \|$ and both of these are thanks
to the Hadamard three-line theorem.

The R\'{e}nyi divergence is simply:
\begin{equation}
D_p(\psi|_{\mathcal{A}}) = \frac{2p}{p-2} \ln \| \psi \|_{L_p(\mathcal{A}')}
\end{equation}
and the notation $D_p(\rho|_{\mathcal{A}}) = D_p(\xi_\rho|_{\mathcal{A}})$ is appropriate. 

\begin{lemma}
\label{lem:Dpbasic}
Fix $2 < p \leq \infty$ and consider some nonzero $\psi \in L_p(\mathcal{A}')$. Then:
\begin{enumerate}
\item Positivity: for normalized states $D_p(\psi|_{\mathcal{A}}) \geq 0$ with equality iff $\omega_\psi|_{\mathcal{A}} = \omega_\eta|_{\mathcal{A}}$.
Also $D_p(\lambda \psi) = D_p(\psi) +  \frac{2p}{p-2} \ln \lambda$ for $\lambda > 0$. 
\item Invariance under unitaries: $D_p(u' \psi|_{\mathcal{A}}) =D_p( \psi|_{\mathcal{A}})$ for $u' \in \mathcal{U}(\mathcal{A}')$. 
\item Monotonicity: for normalized $\psi$, $ p' \rightarrow D_{p'}(\psi|_{\mathcal{A}})$ is a continuous non-decreasing map
for $2 \leq p' \leq p$ with $\lim_{p' \rightarrow 2^+} D_{p'}(\psi|_{\mathcal{A}}) = D(\psi|_{\mathcal{A}})$ the regular divergence.
\end{enumerate}

\end{lemma}
\begin{proof}
Standard \cite{Berta:2016vnw}.
\end{proof}

We can define the following H\"{o}lder product: $L_{p}(\mathcal{A}') \times_{\mathcal{A}'} L_{q}(\mathcal{A}')  \rightarrow L_{r}(\mathcal{A}') $ where $1/p+1/q = 1/r \leq 1/2$
via the polar decompositions (see Section~\ref{sec:proofs} for the extension to $1/r \leq 1$):
\begin{equation}
\psi_1 \times_{\mathcal{A}'} \psi_2  %
\in L_r(\mathcal{A}') \quad \leftrightarrow \quad  u_1' \Delta_{\eta,\varphi_1;\mathcal{A}}^{-1/p}  u_2' \Delta_{\eta,\varphi_2;\mathcal{A}}^{-1/q}  \left| \eta \right> 
\end{equation}
For $p=\infty$ with $\psi_1 = a' \eta$  we replace $ u_1' \Delta_{\eta,\varphi_1;\mathcal{A}}^{-1/p} \rightarrow a'$ in the definition above and similarly if $q=\infty$. 
We use a cumbersome subscript on the product $\times_{\mathcal{A}'}$ since we will need this product for $L_p$ spaces defined using subalgebras and the subalgebra structure is not compatible with the definition
of this product. 
This H\"{o}lder product satisfies H\"{o}lder's inequality $\|\psi_1 \times_{\mathcal{A}'} \psi_2 \|_{L_r(\mathcal{A}')} \leq  \| \psi_1 \|_{L_p(\mathcal{A}')}   \| \psi_2 \|_{L_q(\mathcal{A}')} $ 
and is associative $ (\psi_1 \times_{\mathcal{A}'} \psi_2) \times_{\mathcal{A}'} \psi_3 = \psi_1 \times_{\mathcal{A}'} (\psi_2 \times_{\mathcal{A}'} \psi_3)$.
Consider:
\begin{lemma}
\label{lem:duality}
The $L_q$ norm can be computed as:
\begin{equation}
\| \psi \|_{L_q(\mathcal{A}')} = \sup_{\zeta \in L_p(\mathcal{A}')_{ \leq 1}} \|\zeta \times_{\mathcal{A}'} \psi \|_{L_r(\mathcal{A}')}
\end{equation}
where $1/p +1/q=1/r \leq 1/2$.
\end{lemma}
\begin{proof}
H\"{o}lder's inequality bounds the supremum above by $\|\psi\|_{L_q(\mathcal{A}')}$.
For the reverse inequality, consider the polar decomposition $\psi = \theta' \Delta_{\eta,\varphi;\mathcal{A}}^{-1/q} \eta$
and pick $\zeta =  \Delta_{\eta,\varphi;\mathcal{A}}^{-1/p} (\theta')^\dagger \eta \| \varphi \|^{-2/p}$, which gives the value $\| \varphi \|^{-2/p+2/r} =  \| \psi \|_{L_q(\mathcal{A}')}$.
\end{proof}

\subsection{Positive cones and bimodule action}
 
We note that vectors of the form $ \Delta_{\eta,\varphi;\mathcal{A}}^{-1/p} \eta$ form the positive cone $L^+_p(\mathcal{A}') \subset L_p(\mathcal{A}')$, which can also
be defined as the intersection of $L_p(\mathcal{A}')$ with
the  generalized cones $\mathscr{P}^p(\mathcal{A}') = \overline{\Delta_{\eta;\mathcal{A}}^{-1/2p} (\mathcal{A}')^+ \eta} \subset \mathscr{H}$.
There is an isometric involution $J^p_{\mathcal{A}'}$ defined on $L_p$ that fixes the positive cone and is given as:
\begin{equation}
\psi \rightarrow J^p_{\mathcal{A}'}( \psi) =   (\theta')^\dagger \Delta_{\eta, \theta' j_{\mathcal{A}}(\theta') \varphi ;\mathcal{A}}^{-1/p} \eta
= \Delta_{\eta, \varphi;\mathcal{A}}^{-1/p} (\theta')^\dagger \eta  \in L_p %
\end{equation}
for $\psi = \theta' \Delta_{\eta, \varphi;\mathcal{A}}^{-1/p} \eta $.
 Note $\mathscr{P}^\natural(\mathcal{A}')
= \mathscr{P}^2(\mathcal{A}') =\mathscr{P}^2(\mathcal{A}) $ and $J^{p=2}_{\mathcal{A}'} = J_{\mathcal{A}'} = J_{\mathcal{A}}$.
The notation $\psi \rightarrow \psi^\dagger$ is appropriate; however it suffers from the issue that $\dagger$ depends on the algebra $\mathcal{A}$ and $p$,
so we will mostly avoid this notation.

\begin{lemma}
\label{lem:Jp}
Here we list miscellaneous identities and results:
\begin{enumerate}
\item $\| J^p_{\mathcal{A}'}(\psi) \|_{L_p(\mathcal{A}')} = \| \psi \|_{L_p(\mathcal{A}')}$ and $(J^p_{\mathcal{A}'})^2 = 1$.
\item  $J^p_{\mathcal{A}'}$ acts on $\mathscr{H}$ as an antilinear unbounded operator $= J_{\mathcal{A}} \Delta_{\eta;\mathcal{A}}^{1/p-1/2}$ 
with $L_p(\mathcal{A}')$ contained in the domain of this operator.
\item For a bounded operator $x \in \mathcal{A}'$:
\begin{equation}
  J^p_{\mathcal{A}'} x \Delta_{\eta,\varphi;\mathcal{A}}^{-1/p} \left| \eta \right> =   \Delta_{\eta,\varphi;\mathcal{A}}^{-1/p} x^\dagger \left| \eta \right>
\end{equation}
\item The action of $J^p$ on the H\"{o}lder product satisfies: 
\begin{equation}J^p_{\mathcal{A}'}(\zeta_1 \times_{\mathcal{A}'} \zeta_2 ) = J^{r_2}_{\mathcal{A}'} (\zeta_2) \times_{\mathcal{A}'} J^{r_1}_{\mathcal{A}'} (\zeta_1)
\end{equation}
where $\zeta_i \in L_{r_i}(\mathcal{A}')$ and $1/r_1+1/r_2 = 1/p$.
\end{enumerate}
\end{lemma}
\begin{proof}
(1) Is obvious from the definition.
(2,3) From modular theory the domain condition in (2) is clear; we just need to check the actions agree. We consider the generalized polar decomposition in (3)
and apply an analytically continued cocycle relation, justified by checking the domain conditions of the unbounded modular operators:
\begin{align}
J_{\mathcal{A}} \Delta_{\eta;\mathcal{A}}^{-1/2+1/p} x \Delta_{\eta,\varphi;\mathcal{A}}^{-1/p} \left| \eta \right>    
= J_{\mathcal{A}} \Delta_{\varphi,\eta;\mathcal{A}}^{-1/2+1/p} x \left| \varphi \right>=  \Delta_{\eta,\varphi;\mathcal{A}}^{-1/p}   x^\dagger \left| \eta \right> 
\end{align} 
where we also used various identities for relative modular operators, see for example \cite{araki1982positive,Witten:2018lha}.
Setting $x = \theta'$ we see that $J^p_{\mathcal{A}'} = J_{\mathcal{A}} \Delta_{\eta;\mathcal{A}}^{1/p-1/2}$ thus proving (2) and also proving the identity in (3)
for general $x$ since it is clear that $x \Delta_{\eta,\varphi;\mathcal{A}}^{-1/p} \eta \in L_p(\mathcal{A}')$. (4) We check this identity on $\zeta_{1} = x_{1} \eta$
and $\zeta_2 = J^{r_2}_{\mathcal{A}'} x_2 \eta$ where the right-hand side is $x_2 \Delta_{\eta;\mathcal{A}}^{-1/p} x_1^\dagger \eta$ (justified using the compatibility
of the left action with the H\"{o}lder product, see Lemma~\ref{lem:bimodule}). The left-hand side agrees with this as is most easily checked for $x_1$ in the Tomita algebra
of $\eta$ modular flow so that $\Delta_{\eta}^{1/p} x_1 \Delta_{\eta}^{-1/p}$ is a bounded operator. A density argument along with H\"{o}lder's inequality establishes
the general identity. 
\end{proof}

Now consider $j^p_{\mathcal{A}'}(x) =J^p_{\mathcal{A}'} x J^p_{\mathcal{A}'} $ for $x \in \mathcal{A}'$. This makes
sense as an operator on $L_p(\mathcal{A}')$:
\begin{equation}
j^p_{\mathcal{A}'}(x) : L_p(\mathcal{A}') \rightarrow L_p(\mathcal{A}')  \qquad \forall x \in \mathcal{A}'
\end{equation}
and from Lemma~\ref{lem:Jp} acts to the right of $\psi$:
\begin{equation}
 j^p_{\mathcal{A}'}(x)\theta' \Delta_{\eta,\varphi;\mathcal{A}}^{-1/p} \left| \eta \right>
= \theta' \Delta_{\eta,\varphi;\mathcal{A}}^{-1/p} x^\dagger \left| \eta \right>
\end{equation}
We can also simply act to the left with $y \in \mathcal{A}'$ via the action on $\mathscr{H}$, $y : L_p(\mathcal{A}')  \rightarrow L_p(\mathcal{A}') $.
These two actions commute and give:
\begin{lemma} 
\label{lem:bimodule}
$L_p(\mathcal{A}')$ forms an $\mathcal{A}'$-bimodule with continuous left and right actions given as:
\begin{equation}
\psi \rightarrow y \psi x \equiv y  j^p_{\mathcal{A}'}(x^\dagger) \psi
\end{equation}
for $y,x \in \mathcal{A}'$, forming, respectively, representations of $\mathcal{A}'$ and its opposite algebra. 
We have the estimate $\| y \psi x \|_{L_p(\mathcal{A}')} \leq \| y \| \| x \| \| \psi \|_{L_p(\mathcal{A}')}$.
The action is compatible with the H\"{o}lder product $y (\psi_1 \times_{\mathcal{A}'} \psi_2) x
= (y \psi_1 \times_{\mathcal{A}'} \psi_2 x) = y \times_{\mathcal{A}'} \psi_1 \times_{\mathcal{A}'} \psi_2 \times_{\mathcal{A}'} x $.
\end{lemma}
\begin{proof}
Mostly this is obvious. Continuity of the left and right actions follows from the estimate: 
\begin{align}
\| y \psi x \|_{L_p(\mathcal{A}')} &\leq \| y \| \| \psi x \|_{L_p(\mathcal{A}')}
= \| y \| \| x^\dagger J^p_{\mathcal{A}'}(\psi)  \|_{L_p(\mathcal{A}')}\\& \leq \| y \| \| x \|  \|  J^p_{\mathcal{A}'}(\psi)  \|_{L_p(\mathcal{A}')}
= \| y \| \| x \|  \|\psi  \|_{L_p(\mathcal{A}')}
\end{align}
Compatibility of the left action with the H\"{o}lder product can be checked for a dense subset of $L_{p_2}(\mathcal{A}')$ where $\psi_2 = x_2 \eta$
so that:
\begin{align}
y (\psi_1 \times_{\mathcal{A}'} \psi_2)
& \quad \leftrightarrow  \quad y \theta_1' \Delta_{\eta,\varphi_1}^{-1/p_1} x_2 \eta
= j^{p_1}_{\mathcal{A}'}(x_2^\dagger) y \theta_1' \Delta_{\eta,\varphi_1}^{-1/p_1} \eta
\\ &=  j^{p_1}_{\mathcal{A}'}(x_2^\dagger) \widetilde{\theta_1}' \Delta_{\eta,\widetilde{\varphi_1}}^{-1/p_1} \eta
= \widetilde{\theta_1}' \Delta_{\eta,\widetilde{\varphi_1}}^{-1/p_1}  x_2 \eta
\quad \leftrightarrow \quad (y \psi_1) \times_{\mathcal{A}'} \psi_2
\end{align}
where $(\widetilde{\theta_1}' ,\widetilde{\varphi_1})$ determine the polar decomposition of $y \psi_1$. Density and H\"{o}lder's inequality establish the general
compatibility of the left action. 
Compatibility of the right action can be checked similarly. The third product identity in the chain is definitionally equal to the first. 
\end{proof}

We note that the norm is invariant under left and right unitary actions:
\begin{equation}
\|  u_2 \psi u_1  \|_{L_p} = \| \psi \|_{L_p} \qquad u_1,u_2 \in \mathcal{U}(\mathcal{A}')
\end{equation}
(or partial isometries with appropriate supports).  

It is well known that the modular flows of two different states are unitarily equivalent up to the action of two local unitaries in $\mathcal{A},\mathcal{A}'$.
These unitaries are the Connes cocycles.
The following identities generalize this idea to the action of $\eta$ modular flow on a vector in $L_p$:
\begin{lemma}
\label{lem:modflow}
The action of $\eta$ modular flow isometrically preserves $L_p(\mathcal{A}')$ and splits as follows:
\begin{equation}
\psi \in L_p(\mathcal{A}')  \rightarrow  \Delta_{\eta;\mathcal{A}}^{is} (\psi) = u_s' \psi  j_{\mathcal{A}}(u_{-s})^\dagger %
\end{equation}
where:
\begin{equation}
u_s' = \Delta_{\eta;\mathcal{A}}^{is} \theta' \Delta_{\eta,\varphi;\mathcal{A}}^{-is} \theta'^\dagger
\qquad j_{\mathcal{A}}(u_{-s})  = \Delta_{\eta;\mathcal{A}}^{is} \Delta_{\eta,\varphi;\mathcal{A}}^{-is}
\end{equation}

Similarly the antilinear conjugation splits:
\begin{equation}
J^p_{\mathcal{A}'} (\psi) = \theta'^\dagger \psi (\theta')^\dagger
\end{equation}
where $\theta',\varphi$ determine the polar decomposition of $\psi$ in $L_p$. 
\end{lemma}
\begin{proof}
Straightforward. 
\end{proof}

The identities in Lemma~\ref{lem:modflow} at $p=2$ were very important in the proof of the QNEC. However here they do not play an important role.
We mostly include them here for future reference. 

We need the following result on continuity of the polar decomposition due to \cite{Kosaki:1984uniform}:
\begin{lemma}
\label{lem:polarcont}
Fix $2 \leq p < \infty$ and suppose that $\psi_i \rightarrow \psi$ in $L_p(\mathcal{A}')$. Then the corresponding polar decompositions satisfy:
\begin{equation}
\varphi_i \rightarrow \varphi \qquad so-\lim_i \theta_i' \pi_\varphi(\mathcal{A}') =\theta'
\end{equation}
\end{lemma}
\begin{proof}
We use the isometric identification of $L_p(\mathcal{A}')$ with Haagerup's $L^p$ space (see \cite{araki1982positive,kosaki1984applications}), %
under which the polar decomposition $\psi = \theta' \Delta_{\eta,\varphi;\mathcal{A}}^{-1/p} \eta$ is the polar decomposition of an element of $L^p$:
its absolute value is $|\psi| \equiv | \psi^\dagger \psi|^{1/2} = \Delta_{\eta,\varphi;\mathcal{A}}^{-1/p} \eta \in L_p^+(\mathcal{A}')$ and its phase is $\theta'$.
Then $|\psi|^p \in L_1^+$ corresponds to the positive normal functional represented by $\varphi$, and $|\psi|^{p/2} \in L_2^+$ to the natural cone vector $\varphi = \Delta_{\eta,\varphi;\mathcal{A}}^{-1/2} \eta$.
We need three facts, valid for $1 < p < \infty$ and for general von Neumann algebras:
\begin{enumerate}
\item The absolute value map $\psi \rightarrow |\psi|$ is norm continuous on $L^p$ \cite{Kosaki:1984uniform}. %
\item The map $h \rightarrow h^p$ is a homeomorphism from $L^p_+$ onto $L^1_+$ in the norm topologies (Theorem 4.2 of \cite{Kosaki:1984uniform}). %
\item The natural cone representatives satisfy $\| \xi_\omega - \xi_{\omega'} \|^2 \leq \| \omega - \omega' \|$ \cite{araki1974some}. %
\end{enumerate}
By (1), $|\psi_i| \rightarrow |\psi|$ in $L_p(\mathcal{A}')$. By (2), the corresponding functionals converge in norm, and then by (3) $\varphi_i \rightarrow \varphi$ in $\mathscr{H}$.

For the phases write
\begin{equation}
\theta_i' |\psi| - \theta' |\psi| = \theta_i' \left( |\psi| - |\psi_i| \right) + \left( \psi_i - \psi \right)
\end{equation}
which tends to zero in $L_p(\mathcal{A}')$, since left multiplication by the partial isometry $\theta_i'$ is a contraction (Lemma~\ref{lem:bimodule}).
By H\"{o}lder's inequality and the compatibility of the left action with the H\"{o}lder product (Lemma~\ref{lem:bimodule}), $\theta_i' ( |\psi| \times_{\mathcal{A}'} \zeta) \rightarrow \theta' (|\psi| \times_{\mathcal{A}'} \zeta)$ in $\mathscr{H}$ for all $\zeta \in L_r(\mathcal{A}')$ with $1/p + 1/r = 1/2$.
These vectors are dense in $\pi_\varphi(\mathcal{A}') \mathscr{H}$. So it follows that $so-\lim_i \theta_i' \pi_\varphi(\mathcal{A}') = \theta' \pi_\varphi(\mathcal{A}') = \theta'$.
\end{proof}

\subsection{Subalgebras and \texorpdfstring{$L_p$}{Lp} spaces}

Consider $\mathcal{B} \subset \mathcal{A}$ with a common cyclic and separating $\eta$. 
We have $L_p(\mathcal{A}') \subset L_p(\mathcal{B}')$ which follows from monotonicity of the relative modular operators \cite{Witten:2018lha}.
In particular:
\begin{equation}
\| \psi \|_{L_p(\mathcal{B}')} \leq\| \psi \|_{L_p(\mathcal{A}')} \qquad \psi \in L_p(\mathcal{A}') 
\end{equation}
It follows that $\psi$ has different polar decompositions with $\psi = \theta'_{\mathcal{B}} \Delta_{\eta,\varphi_{\mathcal{B}};\mathcal{B}}^{-1/p} \eta
= \theta' \Delta_{\eta,\varphi;\mathcal{A}}^{-1/p} \eta $. Since the product $\times_{\mathcal{A}'}$ is based on this polar decomposition we have to distinguish
it from $\times_{\mathcal{B}'}$.
We can also define a contraction between these spaces:
\begin{equation}
 J_{\mathcal{B}'}^p J_{\mathcal{A}'}^p  : L_p(\mathcal{A}') \rightarrow L_p(\mathcal{B}') %
\end{equation} 

In fact there is a complex family of such contractions:
\begin{lemma}
\label{lem:Tz}
Fix some $ 2 \leq p < \infty$. The following analytic family of contractions on $\mathscr{H}$ restricts to a contraction between $L_p$ spaces:
\begin{equation}
 T_z(\mathcal{A}' \subset \mathcal{B}') =  \Delta_{\eta;\mathcal{B}}^{-z} \Delta_{\eta;\mathcal{A}}^{z}: L_{p}(\mathcal{A}') \rightarrow L_{p}(\mathcal{B}')\,
\end{equation}
for $0 \leq {\rm Re} z \leq 1/p$. Also $T_{1/p}(\mathcal{A}' \subset \mathcal{B}') = J_{\mathcal{B}'}^p J_{\mathcal{A}'}^p$.
\end{lemma}
\begin{proof}
We first check this at $z=1/p$ acting on $\psi = \Delta_{\eta;\mathcal{A}}^{-1/p} a' \eta$ so that:
\begin{equation}
T_{1/p}(\mathcal{A}' \subset \mathcal{B}') \psi = 
 \Delta_{\eta;\mathcal{B}}^{-1/p}  a' \eta = J_{\mathcal{B}'}^p J_{\mathcal{A}'}^p \psi
\end{equation}
since $\mathcal{A}' \subset \mathcal{B}'$. Density establishes the result. For general $z$ consider:
\begin{align}
\left| \Gamma(z) \right> = \Delta_{\eta,\chi;\mathcal{B}}^{1/p-1/2} \Delta_{\eta;\mathcal{B}}^{-z} \Delta_{\eta;\mathcal{A}}^{z}
\psi
\end{align}
for $\| \chi \| \leq 1$.
A standard argument shows that this has an
 analytic continuation as a vector on $\mathscr{H}$ into the strip $0 \leq {\rm Re} z \leq 1/p$.
 The argument involves checking this statement weakly on $(b' \eta, \cdot)$ and then estimating
 the norm at the edges:
 \begin{align}
 \| \left| \Gamma(1/p+is) \right>  \| = 
 \| \Delta_{\eta,\chi;\mathcal{B}}^{1/p-1/2} \Delta_{\eta;\mathcal{B}}^{-is}  J_{\mathcal{B}'}^p J_{\mathcal{A}'}^p \Delta_{\eta;\mathcal{A}}^{is }
\psi \| \leq \| \psi \|_{L_p(\mathcal{A}')} \\ %
 \| \left| \Gamma(is) \right>  \| = 
 \| \Delta_{\eta,\chi;\mathcal{B}}^{1/p-1/2} \Delta_{\eta;\mathcal{B}}^{-is}  \Delta_{\eta;\mathcal{A}}^{is }
\psi \| \leq \| \psi \|_{L_p(\mathcal{A}')} %
 \end{align}
 and then using the Phragm\'{e}n-Lindel\"{o}f theorem to extend this bound into the interior of the strip.
But:
\begin{equation}
\| T_{z}(\mathcal{A}' \subset \mathcal{B}') \psi \|_{L_p(\mathcal{B}')}
= \sup_{\chi \in \mathscr{H}_{\leq 1}} \| \left| \Gamma(z) \right> \| \leq \| \psi \|_{L_p(\mathcal{A}')}
\end{equation}

\end{proof}

The incompatibility of the product with the subalgebra structure is expressed through the following result:
\begin{lemma}
\label{lem:productincompat}
Consider $\psi, \zeta \in L_{p,r}(\mathcal{A}') \subset L_{p,r}(\mathcal{B}') $ respectively where $1/r+1/p \leq 1/2$ then:
\begin{equation}
  \label{eq:therelation}
 T_{1/p}(\mathcal{A}' \subset \mathcal{B}') \left( J_{\mathcal{A}'}^p(\psi) \times_{\mathcal{A}'} \zeta \right)
= J_{\mathcal{B}'}^p(\psi) \times_{\mathcal{B}'} \zeta
\end{equation}
where the second product $\times_{\mathcal{B}'}$ makes sense since $ \psi, \zeta \in L_{p,r}(\mathcal{B}')$ respectively. 
Here $ T_{1/p}(\mathcal{A}' \subset \mathcal{B}') $ should be understood as a bounded operator on $\mathscr{H}$.
And also for $\psi \in L_p(\mathcal{B}') $ and $\zeta \in L_{r}(\mathcal{A}')$ where $1/r+1/p=1/2$ then:
\begin{equation}
\zeta \times_{\mathcal{A}'}  J_{\mathcal{A}}  T_{1/p}(\mathcal{B} \subset \mathcal{A}) 
J_{\mathcal{B}} \psi
= \zeta \times_{\mathcal{B}'}  \psi 
\end{equation}

\end{lemma}
\begin{proof}
We check the first relation on $\zeta = a' \eta$ which is dense in $L_r(\mathcal{A}')$. The left-hand side:
\begin{equation}
 J_{\mathcal{B}'}^p J_{\mathcal{A}'}^p
 \Delta_{\eta,\varphi;\mathcal{A}}^{-1/p} (\theta')^\dagger a' \eta
 =  J_{\mathcal{B}'}^p (a')^\dagger \psi
 =  J_{\mathcal{B}'}^p (a')^\dagger \theta_{\mathcal{B}}' \Delta_{\eta,\varphi_{\mathcal{B}};\mathcal{B}}^{-1/p} \eta
 = \Delta_{\eta,\varphi_{\mathcal{B}};\mathcal{B}}^{-1/p} ( \theta_{\mathcal{B}}' )^\dagger a' \eta
\end{equation}
and this is clearly the right-hand side.

For the second relation we check this on $\psi = \Delta_{\eta;\mathcal{B}}^{-1/p} b' \eta$ where the left-hand side is:
\begin{align}
\zeta \times_{\mathcal{A}'} \Delta_{\eta;\mathcal{A}}^{1/2-1/p} j_{\mathcal{B}}(b')^\dagger \eta &=
u_{\mathcal{A}}' \Delta_{\eta,\xi_{\mathcal{A}};\mathcal{A}}^{-1/r}\Delta_{\eta;\mathcal{A}}^{1/r} j_{\mathcal{B}}(b')^\dagger \eta   \\
& = j_{\mathcal{B}}(b')^\dagger  u_{\mathcal{B}}' \Delta_{\eta,\xi_{\mathcal{B}};\mathcal{B}}^{-1/r} \eta= \zeta \times_{\mathcal{B}'} \Delta_{\eta;\mathcal{B}}^{1/2-1/p}  j_{\mathcal{B}}(b')^\dagger  \eta
\end{align}
which clearly equals the right-hand side.
\end{proof}
\begin{remark}
\label{rem:therelation}
The relation \eqref{eq:therelation} extends to $1/q = 1/r+1/p \leq 1$, allowing for $L_q$ spaces with $1 < q < \infty$ (see Section~\ref{sec:proofs}).
We will need this later in the proof of Lemma~\ref{lem:finalderiv}, and the proof is essentially the same as the one given above: one checks the relation on the dense set $\zeta = a'\eta$ and extends by continuity, once $T_{1/p}$ is known to be bounded on $L_q$ (for $q<2$ this follows from Lemma~\ref{lem:transpose} in the HSMI case). 
\end{remark}

Now consider the half-sided modular inclusion case \cite{borchers1992cpt,wiesbrock1993half,araki2005extension} where $\mathcal{A}_V = U(-V) \mathcal{A} U(V) \subset \mathcal{A}$
for $V > 0$ and $U(V) = \exp(- i V P)$, with $\Delta_{\eta;\mathcal{A}}^{is} U(V) \Delta_{\eta;\mathcal{A}}^{-is} = U(e^{-2\pi s} V)$ and $J_{\mathcal{A}} U(V) J_{\mathcal{A}} = U(-V)$. We collect here some formulas for the contractions in this case. 
In the half-sided modular inclusion case we can express the $L_p$ norms along $V$ simply by shifting the state with the unitary translation generator. 
\begin{equation}
\| \psi \|_{L_p(\mathcal{A}_V')} =  \| U(V) \psi \|_{L_p(\mathcal{A}')} \leq  \| \psi \|_{L_p(\mathcal{A}')}
\end{equation}
We can also consider:
\begin{equation}
\| J^p_{\mathcal{A}'} U( V) J^p_{\mathcal{A}'} \psi \|_{L_p(\mathcal{A}')} =\| U(V e^{-2\pi i/p} ) \psi  \|_{L_p(\mathcal{A}')} \leq \| \psi \|_{L_p(\mathcal{A}')}
\end{equation}
More generally we have the analytic continuation
\begin{equation}
\label{eq:wedge}
\| U(\alpha) \psi \|_{L_p(\mathcal{A}')} \leq \|\psi \|_{L_p(\mathcal{A}')}
\end{equation}
for $\alpha = V + V_p  e^{-2\pi i/p}$ with $V,V_p \geq 0$. 
In fact:
\begin{equation}
(V,V_p) \rightarrow \| U(\alpha) \psi \|_{L_p(\mathcal{A}')}
\end{equation}
is monotonically non-increasing in both $V$ and $V_p$. 

In terms of the previously defined contraction we have:
\begin{equation}
U(\alpha) = U(V) T_z(\mathcal{A}' \subset \mathcal{A}_V') = U(V e^{-2\pi i z}) \qquad 0 \leq {\rm Re}(z) \leq 1/p
\end{equation}
covering the entire wedge region $V\geq 0, V_p \geq 0$. 

We have the symmetry:
\begin{equation}
J^p_{\mathcal{A}'}  U(\alpha ) J^p_{\mathcal{A}'}  =  U(\alpha^\star e^{-2\pi i/p} )
\end{equation}
which corresponds to $(V,V_p) \rightarrow (V_p,V)$. The positivity preserving maps correspond to $V=V_p$
with $U( \epsilon e^{-\pi i/p} ) : \mathscr{P}^p(\mathcal{A}') \rightarrow  \mathscr{P}^p(\mathcal{A}')$.

The following result gives the support projection on the subalgebra, in the HSMI setting:
\begin{lemma}
\label{lem:support}
Consider some $\psi \in L_p(\mathcal{A}')$ for $2 < p \leq \infty$. Then the support projection satisfies:
\begin{equation}
\pi_\psi(\mathcal{A}_V') = \pi_\psi(\mathcal{A}') \qquad \forall \,\, V \geq 0
\end{equation}
\end{lemma}
\begin{proof}
We use an argument similar to Reeh-Schlieder. We know that $\pi_\psi(\mathcal{A}_V') \leq \pi_\psi(\mathcal{A}')$
and suppose there is a nonzero $\chi \in (\pi_\psi(\mathcal{A}')-\pi_\psi(\mathcal{A}_V'))\mathscr{H}$. Then:
\begin{equation}
\left< \chi \right| U(-a) A U(a) \left| \psi \right> =0 \qquad \forall a \geq V
\end{equation}
for all $ A \in \mathcal{A}$. Consider $\zeta = \Delta_{\eta;\mathcal{A}}^{1/p-1/2} j_{\mathcal{A}}(\alpha^\dagger) \eta \in L_r(\mathcal{A}')$, with $1/r = 1/2-1/p$,
inserted into $J_{\mathcal{A}}( J^r_{\mathcal{A}'}(\zeta) \times_{\mathcal{A}'} J^p_{\mathcal{A}'}(\psi) )
= \psi \times_{\mathcal{A}'} \zeta$ for $\alpha \in \mathcal{A}$ which becomes $\alpha \psi = \psi \times_{\mathcal{A}'} \Delta_{\eta}^{1/p-1/2} j_{\mathcal{A}}(\alpha^\dagger) \eta$.
 For $\alpha = U(-a) A U(a)$ this becomes:
\begin{equation}
\left( \left| \chi \right>, \psi \times_{\mathcal{A}'} U(-a e^{2\pi i/p}) \Delta_{\eta;\mathcal{A}}^{1/p-1/2}  j_\mathcal{A}(A^\dagger) \left| \eta \right>\right)
\end{equation}
Now $U(-a e^{2\pi i/p}) : L_r(\mathcal{A}') \rightarrow L_r(\mathcal{A}')$ has an analytic continuation 
to $U( - a e^{2\pi i/p} y)$ for $0 \leq {\rm Arg }(y) \leq  2\pi(1/2-1/p)$ as a contraction $U( - a e^{2\pi i/p } y):  L_r(\mathcal{A}') \rightarrow L_r(\mathcal{A}')$.
So we have a function of $y$ analytic inside a wedge and continuous on the closure of the wedge. Since it vanishes on the edge for real $y \geq V/a$ it must vanish everywhere.
Hence: 
\begin{equation}
\left< \chi \right| U(-a) A U(a) \left| \psi \right> =0 \qquad a \geq 0
\end{equation}
for all $A \in \mathcal{A}$ and this contradicts $0\neq\chi \in \pi_\psi(\mathcal{A}')\mathscr{H}$ at $a=0$. 
\end{proof}

\section{A Family of Weighted Conditional Norms}
\label{sec:condnorms}

In this section we first work with a von Neumann subalgebra $\mathcal{B} \subset \mathcal{A}$ with a common cyclic and separating $\eta$. 
In this context we will define a new family of weighted conditional norms, generalizing a subfamily of the conditional norms defined
in \cite{Junge:2010mixednorm,Gao:2019subalgebras} for subalgebras with conditional expectations.
Subsequently we specialize to the HSMI inclusion $\mathcal{A}_{V+a} \subset \mathcal{A}_V$ where further results are obtained.

\subsection{General subalgebras}
\label{sec:general}

For $\psi \in L_p(\mathcal{B}') \subset \mathscr{H}$, define the following conditional norm:
\begin{equation}
\label{eq:condnorm}
\| \psi \|_{L_p(\mathcal{A}' \subset \mathcal{B}')}  \equiv \sup_{\zeta \in L_r(\mathcal{A}')_{\leq 1}} \| J^{p}_{\mathcal{B}'}(\psi) \times_{\mathcal{B}'} \zeta \| \qquad \frac{1}{r}+\frac{1}{p} = \frac{1}{2}
\end{equation}
for $2 \leq p \leq \infty$. The product $\times_{\mathcal{B}'}$ makes sense since $L_r(\mathcal{A}') \subset L_r(\mathcal{B}')$. The $J^p_{\mathcal{B}'}$ action 
is inserted as a convenient choice - without $J^p_{\mathcal{B}'}$ the resulting norm is certainly interesting, but mathematically more difficult to work with and hence
less obviously useful to our goal of proving the QNEC.  This is a norm since it can be easily shown to satisfy the triangle inequality, homogeneity and non-degeneracy.
The vector space $L_p(\mathcal{B}')$ need not be complete with respect to this norm.
 We could complete it to define a new Banach space called the ``weighted conditional $L_p$-space: $L_{p}(\mathcal{A}' \subset \mathcal{B}')$'', 
 although we will not go this route in this paper. %
 In the discussion of the QNEC we will further restrict $\psi$ to lie in the subspace $L_p(\mathcal{A}') \subset L_{p}(\mathcal{B}')$
 so we can compare to the case of the trivial inclusion $\mathcal{A}' \subset \mathcal{A}'$.

 Our first result gives another way to compute this quantity in the special case that $\psi \in L_p(\mathcal{A}')$:
 \begin{lemma}
\label{lem:altformula}
Suppose that $\psi \in L_p(\mathcal{A}')$. Then:
\begin{equation}
\label{eq:altformula}
\| \psi \|_{L_p(\mathcal{A}' \subset \mathcal{B}')}  = \sup_{\zeta \in L_r(\mathcal{A}')_{\leq 1}} \| \left(\Delta_{\mathcal{B}}^{-1/r} \Delta_{\mathcal{A}}^{1/r}\right) ( \zeta \times_{\mathcal{A}'} \psi) \|
\end{equation}
More generally suppose that $\mathcal{B}_1 \subset \mathcal{B}_2 \subset \mathcal{A}$ and that $\psi \in L_p(\mathcal{B}'_2)$ then we can write:
\begin{equation}
\label{eq:altformula2}
\| \psi \|_{L_p(\mathcal{A}' \subset \mathcal{B}_1')}  = \sup_{\zeta \in L_r(\mathcal{A}')_{\leq 1}} \| \left(\Delta_{\mathcal{B}_1}^{-1/p} \Delta_{\mathcal{B}_2}^{1/p} \right)
( J^p_{\mathcal{B}_2'} (\psi) \times_{\mathcal{B}_2'} \zeta ) \|
\end{equation}
 \end{lemma}
 \begin{proof}
 Follows from Lemma~\ref{lem:productincompat}. For \eqref{eq:altformula} apply \eqref{eq:therelation} with the roles of $p$ and $r$ exchanged, $T_{1/r}(J^r_{\mathcal{A}'}(\zeta) \times_{\mathcal{A}'} \psi) = J^r_{\mathcal{B}'}(\zeta) \times_{\mathcal{B}'} \psi = J_{\mathcal{B}}(J^p_{\mathcal{B}'}(\psi) \times_{\mathcal{B}'} \zeta)$ (using (4) of Lemma~\ref{lem:Jp}), and note that $J^r_{\mathcal{A}'}$ is an isometric bijection of $L_r(\mathcal{A}')_{\leq 1}$. For \eqref{eq:altformula2} apply \eqref{eq:therelation} to the inclusion $\mathcal{B}_1 \subset \mathcal{B}_2$. 
 \end{proof}
 
 We also have:
 \begin{lemma}
 \label{lem:restricted}
 The variation in \eqref{eq:condnorm} can be restricted so that $\pi_\zeta(\mathcal{A}') \leq \pi_\psi(\mathcal{A}')$
 and using the polar decomposition $\psi = \theta' \Delta_{\eta,\phi;\mathcal{B}}^{-1/p}\eta$ 
 for $\theta' \in \mathcal{B}'$ we can give the explicit formula:
 \begin{equation}
 \| \psi \|_{L_p(\mathcal{A}' \subset \mathcal{B}')} = \sup_{\chi \in \mathscr{H}_{\leq 1}} \sup_{a' \in X } \|   \Delta_{\eta,\phi;\mathcal{B}}^{-1/p} (\theta' )^\dagger (a')^\dagger \Delta_{\eta,\chi;\mathcal{A}}^{1/p-1/2}  \eta \|
 \end{equation}
where the set $X$ could be one of $\mathcal{V}_\psi(\mathcal{A}')\cap \mathcal{V}_\chi(\mathcal{A}')^\dagger ,\mathcal{V}_\psi(\mathcal{A}'),\mathcal{A}'_{\leq 1}$ or $\mathcal{U}(\mathcal{A}')$.
 \end{lemma}
 \begin{proof}
 We note that we have equality of these sets $(\pi_\psi(\mathcal{A}') L_r(\mathcal{A}'))_{\leq 1} = \pi_\psi(\mathcal{A}') L_r(\mathcal{A}')_{\leq 1}
 = \{ \zeta : \pi_\zeta(\mathcal{A}') \leq \pi_\psi(\mathcal{A}'), \| \zeta \|_{L_r(\mathcal{A}')} \leq 1 \}$.
 Under restriction we have:
\begin{align}
&\| \psi \|_{L_p(\mathcal{A}' \subset \mathcal{B}')}  \leq \sup_{\zeta \in L_r(\mathcal{A}')} \frac{\| J^p_{\mathcal{B}'}(\psi) \times_{\mathcal{B}'} \pi_\psi(\mathcal{A}') \zeta \| }
{\| \pi_\psi(\mathcal{A}') \zeta \|_{L_r(\mathcal{A}')}} \\ & \qquad \qquad= 
\sup_{\zeta' \in \pi_\psi(\mathcal{A}') L_r(\mathcal{A}')  } \frac{\| J^p_{\mathcal{B}'}(\psi) \times_{\mathcal{B}'} \zeta' \| }
{\| \zeta' \|_{L_r(\mathcal{A}')}} \leq  \| \psi \|_{L_p(\mathcal{A}' \subset \mathcal{B}')}
\end{align}
The case $p=2$ can be checked independently. So we can take $2 < p \leq \infty$.
$X =  \mathcal{V}_\psi(\mathcal{A}')\cap \mathcal{V}_\chi(\mathcal{A}')^\dagger $ corresponds to the polar
decomposition so this is true automatically. For $X=\mathcal{V}_\psi(\mathcal{A}')$ we can use the polar decomposition for $\zeta$
as $\zeta = (a')^\dagger \Delta_{\eta,\chi;\mathcal{A}}^{1/p-1/2} \eta$ 
with $\| \chi \| \leq 1$ and where $a'$ is a partial isometry $a' (a')^\dagger = \pi_\chi(\mathcal{A}')$
and $(a')^\dagger (a') \leq \pi_\psi(\mathcal{A}')$ but then the larger
set $a' \in \mathcal{V}_\psi(\mathcal{A}')$ gives something contained
in $L_r(\mathcal{A}')_{\leq 1}$ but which also contains all these polar decompositions
so must equal $\pi_\psi(\mathcal{A}') L_r(\mathcal{A}')_{\leq 1}$. The same applies
to $X=\mathcal{A}'_{\leq 1}$ starting with the original non-projected supremum.
For $X= \mathcal{U}(\mathcal{A}')$ we have containment in $L_r(\mathcal{A}')$ so that:
\begin{equation}
 \| \psi \|_{L_p(\mathcal{A}' \subset \mathcal{B}')} \geq \sup_{\chi \in \mathscr{H}_{\leq 1}, u' \in  \mathcal{U}(\mathcal{A}')
  } \|   \Delta_{\eta,\phi;\mathcal{B}}^{-1/p} (\theta' )^\dagger u' \Delta_{\eta,\chi;\mathcal{A}}^{1/p-1/2}  \eta \|
 \end{equation}
 For the opposite inequality we use convexity of the norm and the fact that 
 the convex sums over unitaries $\sum_i p_i u_i$ are in fact dense in $\mathcal{A}'_{\leq 1}$
 so that the $X = \mathcal{A}'_{\leq 1}$ form gives:
 \begin{align}
&\| \psi \|_{L_p(\mathcal{A}' \subset \mathcal{B}')}  \leq
\sum_i p_i  \|  \Delta_{\eta,\phi;\mathcal{B}}^{-1/p} (\theta' )^\dagger u_i' \Delta_{\eta,\chi;\mathcal{A}}^{1/p-1/2}  \eta \|
\leq  \sup_{\chi \in \mathscr{H}_{\leq 1}, u' \in  \mathcal{U}(\mathcal{A}')
  } \|   \Delta_{\eta,\phi;\mathcal{B}}^{-1/p} (\theta' )^\dagger u' \Delta_{\eta,\chi;\mathcal{A}}^{1/p-1/2}  \eta \|
\end{align}
 \end{proof}
 
We then consider some basic properties:
\begin{lemma}
\label{lem:condbasic}
For $\psi \in L_p(\mathcal{B}')$ the conditional norm satisfies:
\begin{enumerate}
\item Trivial inclusion: $\| \psi \|_{L_p(\mathcal{A}' \subset \mathcal{A}')} = \| \psi \|_{L_p(\mathcal{A}')}$
\item Non-conditional bounds:
\[
\| J^p_{\mathcal{B}'} \psi \| \leq \| \psi \|_{L_p(\mathcal{A}' \subset \mathcal{B}')} \leq \| \psi \|_{L_p(\mathcal{B}')}
\]
and hence $\| \eta \|_{L_p(\mathcal{A}' \subset \mathcal{B}')} = 1$.%
\item Left unitary independence on $\mathcal{A}'$: $\| u' \psi \|_{L_p(\mathcal{A}' \subset \mathcal{B}')} = \| \psi \|_{ L_p(\mathcal{A}' \subset \mathcal{B}')}$ for $u' \in \mathcal{A}'$
a partial isometry with $(1-(u')^\dagger u' )\psi =0$. 
\item Right unitary independence on $\mathcal{B}'$: $\| j^p_{\mathcal{B}'}(u') \psi \|_{L_p(\mathcal{A}' \subset \mathcal{B}')} = \| \psi \|_{ L_p(\mathcal{A}' \subset \mathcal{B}')}$
for $u' \in \mathcal{B}'$
a partial isometry with $(1-(u')^\dagger u' )J^p_{\mathcal{B}'}(\psi) =0$.
\item For $2 \leq p' \leq p$ the map 
$\frac{1}{2} - \frac{1}{p'} \rightarrow\| \psi \|_{L_{p'}(\mathcal{A}' \subset \mathcal{B}')} $ is continuous and log-convex.
\item Special cases for $p$:
\begin{equation}
 \| \psi \|_{L_2(\mathcal{A}' \subset \mathcal{B}')} =  \| \psi \| \qquad  \| \psi \|_{L_\infty(\mathcal{A}' \subset \mathcal{B}')} =   \| \psi \|_{L_\infty(\mathcal{B}')}
\end{equation}
\item ALT type inequality for $ 2 \leq p \leq 4$ and $\psi \in L_{4/(4-p)}(\mathcal{A}')$
\begin{equation}
\| \psi \|_{L_p(\mathcal{A}' \subset \mathcal{B}')}
\leq \sup_{u' \in \mathcal{V}_\psi(\mathcal{A}')} \left\|\Delta_{\mathcal{B}}^{-\frac{(p-2)}{4}} \Delta_{\mathcal{A}}^{\frac{(p-2)}{4}} u' \Delta_{\eta,\psi;\mathcal{A}}^{-\frac{(p-2)}{4}} \psi \right\|^{2/p}
\end{equation}
\item Monotonicity for $\mathcal{B}$: for $\mathcal{B}_1 \subset \mathcal{B}_2 \subset \mathcal{A}$ and $\psi \in L_p(\mathcal{B}_2')$ then:
\begin{equation}
\| \psi \|_{L_p(\mathcal{A}' \subset \mathcal{B}_1')} \leq \| \psi \|_{L_p(\mathcal{A}' \subset \mathcal{B}_2')}
\end{equation}
\item Monotonicity for $\mathcal{A}$: for $\mathcal{B} \subset \mathcal{A}_1 \subset \mathcal{A}_2$  then:
\begin{equation}
\| \psi \|_{L_p(\mathcal{A}_2' \subset \mathcal{B}')} \leq \| \psi \|_{L_p(\mathcal{A}_1' \subset \mathcal{B}')}
\end{equation}
\item Lower semicontinuity in the state: suppose $\rho_i \rightarrow \rho$ with $\xi_{\rho_i},\xi_\rho \in L_p(\mathcal{B}')$ then:
\begin{equation}
\liminf_i \| \rho_i \|_{L_p(\mathcal{A}' \subset \mathcal{B}')} \geq \| \rho \|_{L_p(\mathcal{A}' \subset \mathcal{B}')}
\end{equation}
\end{enumerate}
\end{lemma}
\begin{proof}
(1) Clear from \eqref{eq:condnorm} and Lemma~\ref{lem:duality}. (2) Uses the non-commutative H\"{o}lder's inequality. (3) This follows from $J^p_{\mathcal{B}'}(u'\psi) \times_{\mathcal{B}'} \zeta = J^p_{\mathcal{B}'}(\psi) \times_{\mathcal{B}'} (u')^\dagger \zeta$, absorbing $u'$ into $\zeta \rightarrow (u')^\dagger \zeta$. Since $(u')^\dagger u' \geq \pi' = \pi_{\psi}(\mathcal{A}')$
this does not affect the variation over $\zeta$ which can be restricted to $\pi_\zeta(\mathcal{A}') \leq \pi'$ (Lemma~\ref{lem:restricted}). 
 (4) Clear. (5) %
Using the polar decomposition of $\psi = \theta' \Delta_{\eta,\phi;\mathcal{B}}^{-1/p} \eta$ for $\psi \in L_p(\mathcal{B}')$ one can show
the variation is over $\| \Delta_{\eta;\mathcal{B}}^{1/p-1/p'}  \Delta_{\eta,\phi;\mathcal{B}}^{-1/p} (\theta' )^\dagger  \Delta_{\eta,\chi;\mathcal{A}}^{1/p'-1/2} u' \eta \|$
where $\| \chi \| \leq 1$ and $u' \in \mathcal{A}'$ is a partial isometry with $u' (u')^\dagger = \pi_\chi(\mathcal{A}')$. 
 We are free to shift $u' \rightarrow \Delta_{\eta,\chi;\mathcal{A}}^{it} u' \Delta_{\eta;\mathcal{A}}^{-it}$ for any $t \in \mathbb{R}$. Then:
 \begin{align}
\| \psi \|_{L_{p'}(\mathcal{A}' \subset \mathcal{B}')}  &= %
  \sup_{u', \chi,\xi} \sup_{t \in \mathbb{R}}  |f(t + i/r')| \\ f(s) &=   \left< \xi \right| \Delta_{\eta;\mathcal{B}}^{-1/r - is}   \Delta_{\eta,\phi;\mathcal{B}}^{-1/p} (\theta' )^\dagger \Delta_{\eta,\chi;\mathcal{A}}^{is} u' \left| \eta\right>
\end{align}
where $\xi,\chi \in \mathscr{H}_{\leq 1}$. %
Here $1/r' = 1/2-1/p'$.
$f(s)$ is an analytic function inside the strip from $0 < {\rm Im} s < 1/r$, continuous on the closure. An application of the three-line theorem gives log-convexity. 
Due to convexity we get continuity in $p'$ away from the endpoints. We just need to check at the endpoints. At $p'=2$ we use the non-conditional bounds above and at $p' = p$ consider that convexity implies upper semicontinuity at the endpoint:
\begin{align}
\| \psi \|_{L_{p}(\mathcal{A}' \subset \mathcal{B}')} & \geq \limsup_{p' \rightarrow p^-} \| \psi \|_{L_{p'}(\mathcal{A}' \subset \mathcal{B}')} %
\geq \liminf_{p' \rightarrow p^-} \| \Delta_{\eta;\mathcal{B}}^{1/p-1/p'}  \Delta_{\eta,\phi;\mathcal{B}}^{-1/p} (\theta' )^\dagger  \Delta_{\eta,\chi;\mathcal{A}}^{1/p'-1/2} u' \eta \|
\\ &
=  \|  \Delta_{\eta,\phi;\mathcal{B}}^{-1/p} (\theta' )^\dagger  \Delta_{\eta,\chi;\mathcal{A}}^{1/p-1/2} u' \eta \|
\end{align}
Taking the supremum over $\chi,u'$ shows the limit exists and thus continuity to $p$. 
(6) Obvious. (7) Using the freedom to rephase $\psi \rightarrow w' \psi$ for some partial isometry $w' \in \mathcal{A}'$
we can work with the representative in the natural cone $\xi_\psi \in \mathscr{P}^\natural$. Consider the holomorphic vector
where from Lemma~\ref{lem:restricted} we are free to consider $u' \in \mathcal{V}_{\xi_\psi}(\mathcal{A}')$:
\begin{equation}
\left| \Gamma(z) \right> =
\Delta_{\eta;\mathcal{B}}^{ -a(z) } \Delta_{\eta;\mathcal{A}}^{a(z)} \Delta_{\eta,\chi;\mathcal{A}}^{-b(z)} u' \Delta_{\eta,\xi_\psi;\mathcal{A}}^{-c(z)} \eta
\end{equation} 
where $a(z) = zp/r$ and $b(z) = 1/2-z$ and $c(z) = zp/2$. This vector
is well defined for $0 \leq {\rm Re} z \leq 1/2$: firstly check that $\Delta_{\eta,\chi;\mathcal{A}}^{-b(z)} \Delta_{\eta,\xi_\psi;\mathcal{A}}^{-c(z)} \eta$ is well
defined (see for example \cite{araki1982positive}) where the assumption on $\psi$ guarantees $\eta \in \mathcal{D}(\Delta_{\eta,\xi_\psi;\mathcal{A}}^{-c(z)})$
and a three-line estimate establishes the rest. Then acting with $\Delta_{\eta;\mathcal{B}}^{ -a(z) } \Delta_{\eta;\mathcal{A}}^{a(z)}
= T_{a(z)}$, which is a contraction on $\mathscr{H}$ for $0 \leq {\rm Re}\, a(z) \leq 1/2$ (since $\Delta_{\eta;\mathcal{A}} \leq \Delta_{\eta;\mathcal{B}}$ as quadratic forms), so for $0 \leq {\rm Re} z \leq 1/2$ when $2\leq p \leq 4$, gives a holomorphic vector $\Gamma(z)$ with all the usual
properties. The three-line estimate for this new vector at $z =1/p$ is $\| \left| \Gamma(1/p) \right>\| \leq \sup_t \| \left| \Gamma(it) \right>\|^{1-2/p}
 \sup_t \| \left| \Gamma(1/2+it) \right>\|^{2/p}$
\begin{align}
\| \left| \Gamma(it) \right> \|& \leq \|  \Delta_{\eta,\chi;\mathcal{A}}^{-1/2} u' \Delta_{\eta,\xi_\psi;\mathcal{A}}^{-itp/2} \Delta_{\eta;\mathcal{A}}^{itp/2}  \eta \| \leq 1 \\
\| \left| \Gamma(1/2+it) \right> \| & \leq \sup_{u' \in \mathcal{V}_{\xi_\psi}(\mathcal{A}')} \| \Delta_{\eta;\mathcal{B}}^{ -p/2r } \Delta_{\eta;\mathcal{A}}^{p/2r + it (p/2-1)} \Delta_{\eta,\chi;\mathcal{A}}^{i t} u' \Delta_{\eta,\xi_\psi;\mathcal{A}}^{-it p/2 - p/4}  \eta \|  \\
& = \sup_{u' \in \mathcal{V}_{\psi}(\mathcal{A}')} \| \Delta_{\eta;\mathcal{B}}^{ -p/2r } \Delta_{\eta;\mathcal{A}}^{p/2r} u' \Delta_{\eta,\psi;\mathcal{A}}^{- p/4+1/2}  \psi \| 
\end{align}
where we absorbed the extra $t$-dependent partial isometries into $u'$. 

(8) Uses \eqref{eq:altformula2} along with the fact that $\Delta_{\mathcal{B}_1}^{-1/p} \Delta_{\mathcal{B}_2}^{1/p}$ is a contraction since $\Delta_{\mathcal{B}_2} \leq \Delta_{\mathcal{B}_1}$. (9) Uses the original variational formula and
$L_r(\mathcal{A}'_2)_{\leq 1} \subset L_r(\mathcal{A}'_1)_{\leq 1}$. 

(10) For each fixed $\zeta\in L_r(\mathcal{A}')_{\leq 1}$ the map $\xi \rightarrow J^p_{\mathcal{B}'}(\xi)\times_{\mathcal{B}'}\zeta$ is the restriction of a closed antilinear operator with domain containing $L_p(\mathcal{B}')$, so taking $\chi$ in the dense domain of its adjoint proves the weak limit $  \left|\left< \chi \right| \left.  J^{p}_{\mathcal{B}'}(\xi_{\rho}) \times_{\mathcal{B}'} \zeta \right> \right| = \liminf_i \left|\left< \chi \right| \left.  J^{p}_{\mathcal{B}'}(\xi_{\rho_i}) \times_{\mathcal{B}'} \zeta \right> \right|
\leq \liminf_i \| \rho_i \|_{L_p(\mathcal{A}' \subset \mathcal{B}')}$. We can then extend this bound to all $\chi \in \mathscr{H}_{\leq 1}$ allowing us to compute $\sup_{\zeta \in L_r(\mathcal{A}')_{\leq 1}, \chi \in \mathscr{H}_{\leq 1}}$ showing (10). 
\end{proof}

Now consider the conditional entropy for $2 < p \leq \infty$:
\begin{equation}
\label{eq:Hp}
H_p(\psi|_{\mathcal{A} | \mathcal{B}})
= - \left(\frac{1}{2}-\frac{1}{p}\right)^{-1} \ln \frac{\| \psi \|_{L_{p}(\mathcal{A}' \subset \mathcal{B}')}}
{\| \psi \|_{L_{p}(\mathcal{B}')}}
\end{equation}
As we will see this is invariant under unitaries $u' \in \mathcal{A}'$ so it is actually a function of the state
$\rho \in \mathcal{A}_\star^+$. The following notation is then appropriate:
\begin{equation}
H_p(\rho|_{\mathcal{A} | \mathcal{B}})
\equiv H_p(\xi_\rho |_{\mathcal{A} | \mathcal{B}})
\end{equation}
However we will mostly use the purified state notation. 

Many properties follow directly from Lemma~\ref{lem:condbasic}. We list here some of the most important properties:
\begin{lemma}
\label{lem:Hpbasic}
The weighted conditional entropy satisfies:
\begin{enumerate}
\item Invariance under $\psi \rightarrow u' \psi$ for $u' \in \mathcal{U}(\mathcal{A}')$.
\item Invariance under $\psi \rightarrow j^p_{\mathcal{B}'} \circ j_{\mathcal{B}'}(u) \psi$ for $u \in \mathcal{U}(\mathcal{B})$.
\item Positivity: $ H_p(\psi|_{ \mathcal{A}|\mathcal{B}}) \geq 0$ and $H_p(\psi|_{ \mathcal{A}|\mathcal{A}})= 0$
\item $\mathcal{A}$-monotonicity: $H_p(\psi|_{\mathcal{A}_1|\mathcal{B}}) \leq H_p(\psi|_{\mathcal{A}_2|\mathcal{B}}) $
for $\mathcal{B} \subset \mathcal{A}_1 \subset \mathcal{A}_2$.
\item Continuity in $p'$: $p' \rightarrow H_{p'}(\psi|_{ \mathcal{A}|\mathcal{B}})$ is continuous for $2 < p' \leq p$.
\item Monotonicity in $p'$: $p' \rightarrow H_{p'}(\psi|_{ \mathcal{A}|\mathcal{B}}) - D_{p'}(\psi|_\mathcal{B})$ is monotonically non-increasing for $2 < p' \leq p$ for normalized $\| \psi \| = 1$.

\item Special values:  for $p=\infty$ then $H_\infty(\psi|_{ \mathcal{A}|\mathcal{B}}) = 0$. 
\end{enumerate}
\end{lemma}
\begin{proof}
(1-5) follow directly from the properties proven in Lemma~\ref{lem:condbasic} for the conditional norm.
(6) Note $H_{p'}(\psi|_{\mathcal{A}|\mathcal{B}}) - D_{p'}(\psi|_\mathcal{B}) = -r' \ln \| \psi \|_{L_{p'}(\mathcal{A}' \subset \mathcal{B}')}$ with $1/r' = 1/2-1/p'$. By (5) of Lemma~\ref{lem:condbasic} this logarithm is convex in $1/r'$ and vanishes at $1/r'=0$ (for normalized $\psi$), so $r'$ times it is non-decreasing in $1/r'$.
(7) The case $p=\infty$ requires $\psi = b' \eta$ and $H_\infty =0$ follows from a straightforward calculation.
\end{proof}

The definition we have given does not work at $p=2$. We can constrain the limit:
\begin{lemma}
\label{lem:Hplimit}
Consider $\psi \in L_p(\mathcal{A}')$ with $\|\psi\|=1$ for $2 < p \leq \infty$. Then the limit as $p'\to 2^+$ exists (possibly $=\infty$) and satisfies:
\begin{align}
\label{eq:Hplimitbounds}
&\limsup_{\epsilon \rightarrow 0^+} \inf_{u' \in \mathcal{V}_\psi(\mathcal{A}')}
- \frac{1}{\epsilon} \ln \| \Delta_{\mathcal{B}}^{-\epsilon}  \Delta_{\mathcal{A}}^{\epsilon} u' \psi \|
\\ & \qquad \qquad \leq \lim_{p' \rightarrow 2^+} H_{p'}(\psi|_{\mathcal{A}|\mathcal{B}}) + D(\psi|_\mathcal{A}) - D(\psi|_\mathcal{B})  \leq 
 \inf_{u' \in  \mathcal{V}_\psi(\mathcal{A}')} \liminf_{\epsilon \rightarrow 0^+}
-\frac{1}{\epsilon} \ln \| \Delta_{\mathcal{B}}^{-\epsilon}  \Delta_{\mathcal{A}}^{\epsilon} u' \psi \| \nonumber
\end{align} 
\end{lemma}
\begin{proof}
The left and right bounds come respectively from the ALT bound, (7) of Lemma~\ref{lem:condbasic}, and
picking $\chi = \psi$ in the variational form \eqref{eq:altformula} of the norm, with
$\epsilon = (p-2)/4$ and $\epsilon = 1/2 - 1/p$, respectively. Actually we get the
quoted bounds \eqref{eq:Hplimitbounds} but with the replacement on the left and right:
\begin{equation}
  \psi \rightarrow \psi_\epsilon = \frac{\Delta_{\eta,\psi; \mathcal{A}}^{ - \epsilon}
\psi}{\| \Delta_{\eta,\psi;\mathcal{A}}^{ - \epsilon}
\psi \| }
\end{equation}
and the shift in the normalization yields the extra relative entropy term in the middle:
\begin{equation}
\lim_{\epsilon \rightarrow 0^+} \frac{1}{\epsilon} \ln \| \Delta_{\eta,\psi;\mathcal{A}}^{ - \epsilon}
\psi \| = D(\psi|_{\mathcal{A}}) 
\end{equation}
Our goal is to show that replacing $\psi_\epsilon \rightarrow \psi$ does not affect the bounds.
We use the technical Lemma~\ref{lem:technical}.
For this, let us compute:
\begin{equation}
\| \psi - \psi_\epsilon \|^2 = 2\left( 1 - \frac{{\rm Re} \left< \psi \right|\Delta^{-\epsilon}_{\eta,\psi;\mathcal{A}}\left| \psi \right> }{\| \Delta^{-\epsilon}_{\eta,\psi;\mathcal{A}} \psi \|}\right)
\end{equation}
Now since $\psi \in L_p(\mathcal{A}')$ we know that $\psi \in \mathcal{D}(\Delta^{-\epsilon}_{\eta,\psi;\mathcal{A}})$ for $\epsilon \leq 1/2-1/p$
and $\psi \in\mathcal{D}( \Delta^{\epsilon}_{\eta,\psi;\mathcal{A}})$ for $\epsilon \leq 1/2$ hence the above correlation functions have analytic extensions in the neighborhood
of $\epsilon =0$. This means we can Taylor expand about $\epsilon=0$ with:
\begin{equation}
\| \psi - \psi_\epsilon \|^2
= \epsilon^2 \| \left(\ln \Delta_{\eta,\psi;\mathcal{A}} + D(\psi|{\mathcal{A}}) \right) \psi\|^2
+ \mathcal{O}(\epsilon^3)
\end{equation}
that is, the fluctuations of the relative modular Hamiltonian are finite. 
We then use the technical lemma twice, with $\Sigma = \mathcal{V}_\psi(\mathcal{A}')$ for the ALT side and $\Sigma = \{ u' \}$
with $u' \in \mathcal{V}_\psi(\mathcal{A}')$ on the other side.
This allows us to replace $\psi_\epsilon \rightarrow \psi$ on both sides which gives the claimed bounds.
\end{proof}

We used the following technical lemma:
\begin{lemma}
\label{lem:technical}
Consider some $\psi \in \mathscr{H}$. Given any set of operators $\Sigma \subset \mathcal{B}(\mathscr{H})_{\leq 1}$ containing some $x' \in \Sigma$
such that $(x')^\dagger x' \psi = \psi$, and a parameterized set of %
states $\epsilon \rightarrow \psi_\epsilon$ such that: 
\begin{equation} 
\| \psi_\epsilon \| = \| \psi \| = 1 \, \,,\quad%
\| \psi - \psi_\epsilon \| = \mathcal{O}(\epsilon) \qquad \psi \notin \bigcap_{u' \in \Sigma} {\rm ker}(u')%
\end{equation}
then:
\begin{equation}
\label{eq:technical}
\lim_{\epsilon \rightarrow 0^+} \frac{1}{\epsilon} \left|   \sup_{u' \in \Sigma} \ln \| \Delta_{\mathcal{B}}^{-\epsilon} \Delta_{\mathcal{A}}^{\epsilon} u' \psi_\epsilon \| 
-\sup_{u' \in \Sigma}  \ln \| \Delta_{\mathcal{B}}^{-\epsilon} \Delta_{\mathcal{A}}^{\epsilon} u' \psi \| \right| =0
\end{equation}
\end{lemma}
\begin{proof}
See Appendix~\ref{app:proofs}. 
\end{proof}

Given this it is natural to introduce an explicit expression at $p=2$:
\begin{align}
H_2(\psi|_{\mathcal{A} | \mathcal{B}})
&\equiv\Theta(\psi|_{\mathcal{A} | \mathcal{B}}) \label{eq:H2general}
- D(\psi|_\mathcal{A}) + D(\psi|_\mathcal{B}) \\
\Theta(\psi|_{\mathcal{A} | \mathcal{B}})
 &\equiv \inf_{u' \in \mathcal{V}_\psi(\mathcal{A}')} \left< \psi \right|(u')^\dagger \mathcal{P}(\mathcal{B} \subset \mathcal{A})  u' \left| \psi \right>
 \qquad \mathcal{P}(\mathcal{B} \subset \mathcal{A}) = \overline{ \ln \Delta_{\mathcal{B}} - \ln \Delta_{\mathcal{A}} }
\end{align}
where we have implicitly assumed that  $\ln \Delta_{\mathcal{B}} - \ln \Delta_{\mathcal{A}}$ is essentially self-adjoint.
If this is not the case we could simply interpret the right-hand side of the second line as a positive quadratic form. 
This formula is motivated by the HSMI case, where indeed the self-adjoint operator $ \mathcal{P}(\mathcal{B} \subset \mathcal{A}) $ exists and
where we will show $H_p$ has a continuous limit to $p=2$. For more general inclusions it is not
clear to us whether this form is useful; in general it could be very badly behaved (almost always $=\infty$)
 for example in the case that $\ln \Delta_{\mathcal{B}} - \ln \Delta_{\mathcal{A}}$
is not densely defined or essentially self-adjoint. Even if it is essentially self-adjoint, it has proven difficult for us to show any relation between $\lim_{p \rightarrow 2^+} H_p$ and $H_2$
without the full machinery of the HSMI case, most notably, the log-convexity in (2) of Lemma~\ref{lem:HSMInorm}, which gives the monotonicity in $\epsilon$ used in the proof of Lemma~\ref{lem:HpHSMI}.

\subsection{HSMI}
\label{sec:HSMI}

We now specialize to the case of an HSMI with $\mathcal{A}_a \subset \mathcal{A}$ for $a \geq 0$. 
The norm and conditional entropies satisfy a few
more properties as we vary $a$ continuously and we can say more about the limit to $p=2$. We will consider $\psi \in L_p(\mathcal{A}')$ so all the conditional norms are defined for $a \geq 0$. 
In this case the variational form \eqref{eq:altformula} is:
\begin{equation}
\label{eq:HSMIvar}
\| \psi \|_{L_p(\mathcal{A}' \subset \mathcal{A}_a')}  = \sup_{\chi \in \mathscr{H}_1, u' \in \mathcal{U}(\mathcal{A}')} \| U(a e^{-2\pi i /p}) u' \Delta_{\eta,\chi;\mathcal{A}}^{1/p-1/2} \psi \|
\end{equation}
for $ a\geq 0, 2 \leq p \leq \infty$. 

Many properties follow from the more general case in Lemma~\ref{lem:condbasic}.  
The extra properties that apply for HSMI are as follows:
\begin{lemma}
\label{lem:HSMInorm}
Consider some $\psi \in L_p(\mathcal{A}')$ for $2 \leq p \leq \infty$. Then we have:
\begin{enumerate}
\item Boost covariance: $\| \Delta_{\eta}^{is} \psi \|_{L_p(\mathcal{A}' \subset \mathcal{A}_a')} = \| \psi \|_{L_p(\mathcal{A}' \subset \mathcal{A}_{a e^{2\pi s}}')} $  for all $s \in \mathbb{R}$. 
\item $a \rightarrow \| \psi \|_{L_p(\mathcal{A}' \subset \mathcal{A}_a')}$ is continuous, log-convex and monotonically non-increasing for $0 \leq a < \infty$
\item $b \rightarrow \| \psi \|_{L_p(\mathcal{A}_b' \subset \mathcal{A}_a')}$ is continuous, 
monotonically non-decreasing for $0 \leq b \leq a$.
\item Bound from large $a$, assuming $\eta$ is the unique vacuum state of $P$:
\begin{equation}
  \| \psi \|_{L_p(\mathcal{A}' \subset \mathcal{A}_a')}  \geq %
\| \psi \|_{L_{2p/(2+p)}(\mathcal{A}')} \geq F(\psi|\eta;\mathcal{A}) \qquad 2 < p < \infty
\end{equation}%
\end{enumerate}
\end{lemma}
\begin{proof}
(1) from an explicit calculation, using $\Delta_\eta^{-is} \mathcal{A}_a \Delta_\eta^{is} = \mathcal{A}_{a e^{2\pi s}}$. (2) apply H\"{o}lder's inequality to the integral over the spectral measure of $P$ in the state $\Psi = u' \Delta_{\eta,\chi;\mathcal{A}}^{1/p-1/2} \psi $:
\begin{equation}
\left< \Psi \right| e^{ - 2 a_\lambda \sin(2\pi/p) P} \left| \Psi \right>
\leq (\left< \Psi \right| e^{ - 2 a_1 \sin(2\pi/p) P} \left| \Psi \right>)^\lambda 
(\left< \Psi \right| e^{ - 2 a_2 \sin(2\pi/p) P} \left| \Psi \right>)^{1-\lambda }
\end{equation}
where $a_\lambda = a_1 \lambda + a_2 (1-\lambda)$. This proves that $a \rightarrow \| e^{ - a \sin(2\pi/p) P} \left| \Psi \right>\|$ is log-convex.
The pointwise supremum over a family of log-convex functions is still log-convex. Since $\| \psi \|_{L_p(\mathcal{A}' \subset \mathcal{A}_a')} \leq \| \psi \|_{L_p(\mathcal{A}' \subset \mathcal{A}_a')}|_{a=0} $
combined with convexity we have monotonicity. Continuity follows away from the endpoints. At $a=0$ we use monotonicity so that 
\begin{equation}
\| \psi \|_{L_p(\mathcal{A}')} = \| \psi \|_{L_p(\mathcal{A}' \subset \mathcal{A}_a')}|_{a=0}
\geq \lim_{a \rightarrow 0} \| \psi \|_{L_p(\mathcal{A}' \subset \mathcal{A}_a')} \geq \| \Delta_{\eta,\chi;\mathcal{A}}^{1/p-1/2} \psi \|
\end{equation}
after which taking the supremum over $\chi$ establishes the limit. 
(3) The case $p=\infty$ is clear. Consider $2 \leq p < \infty$. 
We can express the shifted norm as:
\begin{align}
  \label{eq:weakcont}
\| \psi \|_{L_p(\mathcal{A}_b' \subset \mathcal{A}_a')}  &= \sup_{\chi \in \mathscr{H}_1, u' \in \mathcal{U}(\mathcal{A}')} \| U((a-b) e^{-2\pi i /p})  u' \Delta_{\eta,\chi;\mathcal{A}}^{1/p-1/2} U(b) \left|\psi\right> \| \\
& = \| U(b) \psi \|_{L_p(\mathcal{A}' \subset \mathcal{A}'_{a-b})} 
\end{align}
where $U(b) \psi \in L_p(\mathcal{A}')$.
For $b$ approaching $b_0$:
\begin{align}
\left| \| \psi \|_{L_p(\mathcal{A}_b' \subset \mathcal{A}_a')} - \| \psi \|_{L_p(\mathcal{A}_{b_0}' \subset \mathcal{A}_a')} \right|   
& \leq \left| \| U(b) \psi \|_{L_p(\mathcal{A}' \subset \mathcal{A}'_{a-b})} - \| U(b_0) \psi \|_{L_p(\mathcal{A}' \subset \mathcal{A}'_{a-b})} \right| \\ &+
\left| \| U(b_0) \psi \|_{L_p(\mathcal{A}' \subset \mathcal{A}'_{a-b})} - \| U(b_0) \psi \|_{L_p(\mathcal{A}' \subset \mathcal{A}'_{a-b_0})} \right|
\end{align}
where the first term can be estimated using the reverse triangle inequality and the bound: 
\begin{equation}
\leq \| (U(b)- U(b_0)) \left|\psi\right> \|_{L_p(\mathcal{A}' \subset \mathcal{A}_{a-b}')} \leq \| (U(b)- U(b_0)) \left|\psi\right> \|_{L_p(\mathcal{A}')}
\end{equation}
which vanishes as $ b\rightarrow b_0$ by continuity of the $L_p$ norms
$\| U(b) \psi \|_{L_p(\mathcal{A}')} \rightarrow \| U(b_0)\psi \|_{L_p(\mathcal{A}')} $ \footnote{\label{fn:notcircular}This follows from
convexity in $b$ and the QNEC statement, Theorem~\ref{thm:main2} which we are yet to prove; our future proof in Section~\ref{sec:proofs} will not depend on this part (3) of the Lemma.}
plus weak convergence in a uniformly convex Banach space. The second term continuously vanishes as $b \rightarrow b_0$ due to part (2) of this Lemma. 

(4)  The limit $a \rightarrow \infty$ can be achieved by $\inf_a$
and switching this with the $\sup_{\chi,u'}$ at the expense of a $\geq$ allows us to project to the vacuum:
\begin{equation}
\| \psi \|_{L_p(\mathcal{A}' \subset \mathcal{A}_a')}
\geq \sup_{\chi,u'} | \left< \eta \right| u' \Delta_{\eta,\chi;\mathcal{A}}^{1/p-1/2} \left|\psi \right>|
= \sup_{\zeta \in L_r(\mathcal{A}')_{\leq 1}} | \left< \zeta, \psi \right> |
= \| \psi \|_{L_{r'} (\mathcal{A}')} \qquad \frac{1}{r}+ \frac{1}{r'} = 1
\end{equation}
where we used $\psi \in L_p(\mathcal{A}') \subset L_{r'}(\mathcal{A}')$. The dual pairing between
$L_{r}$ and $L_{r'}$ then
gives the result.  
\end{proof}

Consider now the weighted conditional entropy. Most HSMI properties follow directly from properties proven above for the  conditional $L_p$ norms. 
We will also consider the limit to $p=2$ where we need a new definition.
The operator introduced in \eqref{eq:H2general} is $\mathcal{P}(\mathcal{A}_a \subset \mathcal{A}) = 2\pi P a$
so the natural definition for $H_2$ is:
\begin{equation}
  \label{eq:H2def}
H_2(\psi|_{\mathcal{A}|\mathcal{A}_a}) \equiv  2\pi a\, \Theta(\psi|_\mathcal{A}) - D(\psi|_{\mathcal{A}}) +D(\psi|_{\mathcal{A}_a})\,, \quad D(\psi|_{\mathcal{A}}) < \infty
\end{equation}
for $a > 0$ and $H_2(\psi|_{\mathcal{A}|\mathcal{A}}) \equiv 0$, where: 
\begin{equation}
\Theta(\psi|_\mathcal{A}) \equiv \inf_{u' \in \mathcal{V}_\psi(\mathcal{A}')} \left< \psi \right| (u')^\dagger P u' \left| \psi \right> 
\end{equation}
Note that $H_2$ without the condition $D(\psi|_{\mathcal{A}}) < \infty$ would suffer from possible $\infty -\infty$ ambiguities. 
Then, for $a>0$, $H_2(\psi|_{\mathcal{A}|\mathcal{A}_a}) \in [0,\infty]$ with value $\infty$ iff $\Theta(\psi|_\mathcal{A}) = \infty$.  

Note that when finite, $\Theta(\psi|_\mathcal{A}) = \frac{1}{2} \| \sqrt{P} \xi_\psi \|^2$ with $\xi_\psi \in \mathscr{P}^\natural(\mathcal{A})$.
This definition of the quantum expansion is the same as discussed in the introduction
and will agree with the $p \rightarrow 2$ limit of $\Theta_p$ defined later in \eqref{eq:Thetap} of Section~\ref{sec:expansions}, as long as we demand $\psi$ is normalized.
Now in \cite{Faulkner:QNECwithoutAssumptions} (Corollary 1) we derived a sum rule that illuminates this quantity:
\begin{equation}
\label{eq:sumrule}
 D(\psi|_{\mathcal{A}}) +D(\psi|_{\mathcal{A}_a'}) = D(\psi|_{\mathcal{A}_a}) + D(\psi|_{\mathcal{A}'})
 + 2\pi a\max \begin{cases} \Theta(\psi|_\mathcal{A})  + \Theta(\psi|_{\mathcal{A}'}) \\
\Theta(\psi|_{\mathcal{A}_a})  + \Theta(\psi|_{\mathcal{A}_a'})  \end{cases}
\end{equation}
This relation is unambiguous in the presence of $\infty$ values - both sides could be $=\infty$.
It is also useful to recall (Lemma 12 of \cite{Faulkner:QNECwithoutAssumptions}) that if $\psi \in \mathcal{D}(\sqrt{P})$ then
$\Theta(\psi|_{\mathcal{A}_V}) + \Theta(\psi|_{\mathcal{A}_V'}) = \| \sqrt{P} \psi \|^2$ applies for all $V$. 

We can use this to show invariance under $\psi \rightarrow u u' \psi$ for $u' \in \mathcal{A}'$ (obvious)
and $u \in \mathcal{A}_a$ in the following sense:
\begin{lemma}
\label{lem:H2invariance}
Suppose that $D(\psi|_{\mathcal{A}}) < \infty$ and for some partial isometry $u \in \mathcal{V}_\psi(\mathcal{A}_a)$ 
we also have $D(u \psi|_{\mathcal{A}}) < \infty$  then:
\begin{equation}
  \label{eq:requiredrelation}
H_2(\psi|_{\mathcal{A}|\mathcal{A}_a}) = H_2(u  \psi|_{\mathcal{A}|\mathcal{A}_a})
\end{equation}
\end{lemma}
\begin{proof}
The case $a=0$ is immediate from the definition, so assume $a>0$.
We use the purification $\xi = \xi_{\omega_\psi|_{\mathcal{A}}}$ which leaves
invariant $H_2(\psi|_{\mathcal{A}|\mathcal{A}_a})  = H_2(\xi|_{\mathcal{A}|\mathcal{A}_a}) $.
Suppose that $H_2(\xi|_{\mathcal{A}|\mathcal{A}_a}) < \infty$ then 
$\Theta(\xi|_{\mathcal{A}}) < \infty$ so we have finite $D(\xi|_{\mathcal{A}'})
= D(\xi|_{\mathcal{A}})$ and $\xi \in \mathcal{D}(\sqrt{P})$ hence the sum rule applies
with finite quantities:
\begin{equation}
D(\xi|_{\mathcal{A}_a'}) = D(\psi|_{\mathcal{A}_a})  + 4\pi a \Theta(\psi|_{\mathcal{A}})
\end{equation}
For $u \xi$ the sum rule is also finite (all terms are finite) since the left-hand side remains finite by assumption.
We can write the finite sum rule as:
\begin{equation}
  \label{eq:twoexp}
  2\pi a \Theta(u \psi|_{\mathcal{A}}) - D(u \psi|_{\mathcal{A}}) + D(u \psi|_{\mathcal{A}_a}) 
  = - 2\pi a \Theta(\xi|_{\mathcal{A}'})+ D(\xi|_{\mathcal{A}_a'}) - D(\xi|_{\mathcal{A}'})
\end{equation}
Then use $\Theta(u \xi|_{\mathcal{A}'})   = \Theta( \xi|_{\mathcal{A}'})  =
\Theta(\psi|_{\mathcal{A}}) $ and $D(\xi|_{\mathcal{A}'}) = D(\psi|_{\mathcal{A}})$.   This implies $ \Theta(u \psi|_{\mathcal{A}}) < \infty \implies H_2(u  \psi|_{\mathcal{A}|\mathcal{A}_a}) < \infty$ and 
 gives the required relation \eqref{eq:requiredrelation}. 
 Starting by assuming $H_2(u  \psi|_{\mathcal{A}|\mathcal{A}_a}) < \infty$ and running the same argument 
 covers all cases including possible $\infty$ values.   
\end{proof}

For $a>0$, this gives an unambiguous expression for $H_2$ without assuming finite relative entropy:
\begin{equation}
  \label{eq:H2newdef}
H_2(\psi|_{\mathcal{A}|\mathcal{A}_a}) \equiv \begin{cases} 2\pi a\, \Theta( u\psi|_\mathcal{A}) - D(u \psi|_{\mathcal{A}}) +D( u \psi|_{\mathcal{A}_a})\,, & \exists u \in \mathcal{V}_\psi( \mathcal{A}_a)\, : D(u \psi|_{\mathcal{A}})  < \infty\\ \infty\,, & {\rm otherwise} 
 \end{cases}
\end{equation}
where Lemma~\ref{lem:H2invariance} proves that any two finite contributions will give the same value and this is equivalent
to our previous expression where the latter definition is well defined. 
Also note that in this form it is clear that we have unqualified unitary invariance: $H_2( w'w \psi|_{\mathcal{A}|\mathcal{A}_a}) = H_2(\psi|_{\mathcal{A}|\mathcal{A}_a})$ for $w' \in \mathcal{U}(\mathcal{A}')$ and $w \in \mathcal{U}(\mathcal{A}_a)$.

We have the following properties for $H_p$:
\begin{lemma}
\label{lem:HpHSMI}
Fix some $2 \leq p \leq \infty$ and consider $\psi \in L_p(\mathcal{A}')$. Then the following properties
hold for $H_p(\psi|_{\mathcal{A}|\mathcal{A}_a}) \in [0,\infty]$ where we allow for values $\infty$
only at $p=2$:
\begin{enumerate}
\item Invariance under unitaries: $H_p(\psi|_{ \mathcal{A}|\mathcal{A}_a}) $ is invariant under $\psi \rightarrow u' j^p_{\mathcal{A}_a'} ( j_{\mathcal{A}_a}(u)) \psi$
for $u' \in \mathcal{U}(\mathcal{A}')$ and $u \in \mathcal{U}(\mathcal{A}_a)$ including for $p=2$ where $\psi \rightarrow u' u  \psi$.
\item Positivity: $ H_p(\psi|_{ \mathcal{A}|\mathcal{A}_a})  \geq 0$. And  we have equality if $a=0$ or if 
$\psi =  \eta$ or if $\psi$ lies in any of the unitary orbits of $\eta$ implied by (1).
 \item $a$-Continuity: $a \rightarrow H_p(\psi|_{ \mathcal{A}|\mathcal{A}_a})$ is continuous 
 on subsets where it is finite.  It is monotonically non-decreasing at $p=2$. 
 \item $b$-Continuity and monotonicity: $b \rightarrow H_p(\psi|_{ \mathcal{A}_b|\mathcal{A}_a}) $ is
 monotonically non-increasing and continuous for $0 \leq b \leq a$ (where $H_2(\psi|_{ \mathcal{A}_b|\mathcal{A}_a})< \infty$ for $p=2$)
\item Limit to $p' \rightarrow 2$: for $p > 2$ and normalized $\| \psi \|=1$,
\begin{equation}
\lim_{p' \rightarrow 2^+} H_{p'}(\psi|_{\mathcal{A}|\mathcal{A}_a})  = H_2(\psi|_{\mathcal{A}|\mathcal{A}_a}) %
\end{equation}
faithful to the possibly infinite value on the right. 
\end{enumerate}
\end{lemma}
\begin{proof}
(1) For $p=2$, this follows from \eqref{eq:H2newdef} and Lemma~\ref{lem:H2invariance}. 
(2) Clear for $p>2$ and for $p=2$ positivity follows from convexity of $V \rightarrow D(\psi|_V)$. (3) For $p > 2$ this can be found in Lemma~\ref{lem:HSMInorm} along with continuity of the denominator inside
the log $\| \psi \|_{L_p(\mathcal{A}_a')}$ (this will be proven later when we prove Theorem~\ref{thm:main2} but as in Footnote~\ref{fn:notcircular} our proof of this theorem will not hinge on this part of the current lemma).  For $p=2$ it follows from the definition \eqref{eq:H2newdef} and continuity of $D(u \psi|_{\mathcal{A}_a})$.
(4) This is established at $p> 2$ in Lemma~\ref{lem:HSMInorm}. For $p =2$ monotonicity follows
directly from convexity of relative entropy as a function of $b$. Right continuity in $b$ follows from \eqref{eq:H2newdef} 
and the right continuity of $\Theta$ (Lemma 8 of \cite{Faulkner:QNECwithoutAssumptions}). Left continuity follows from
the purified expression \eqref{eq:twoexp} applied with $\mathcal{A} \rightarrow \mathcal{A}_b$
and $\mathcal{A}' \rightarrow \mathcal{A}_b'$ where we use the same fixed natural cone purification $\xi$
associated to $\mathcal{A}$, since $b \rightarrow \Theta(\xi|_{\mathcal{A}_b'})$ is left continuous (Lemma 8 of \cite{Faulkner:QNECwithoutAssumptions} for the commutant family). Hence we have continuity.

(5) The case $a=0$ is immediate, so assume $a>0$. Note that with $\psi \in L_p(\mathcal{A}')$ with $p > 2$ it follows automatically that $D(\psi|_{\mathcal{A}}) < \infty$ so we may utilize the first definition in \eqref{eq:H2def}.
We will use the two-sided estimate \eqref{eq:Hplimitbounds} proven in Lemma~\ref{lem:Hplimit}. In the HSMI case we can now directly take the $\lim_{\epsilon \rightarrow 0}$  on the right-hand side, due
to monotonicity in $\epsilon$ after simplifying $\Delta_{\mathcal{A}_a}^{-\epsilon}  \Delta_{\mathcal{A}}^{\epsilon}  %
\approx e^{ - 2\pi a \epsilon P}$ for small $\epsilon$:
\begin{equation}
\lim_{p' \rightarrow 2^+} H_{p'}(\psi|_{ \mathcal{A}|\mathcal{A}_a}) \leq H_2(\psi|_{ \mathcal{A}|\mathcal{A}_a})
\end{equation}
The left-hand side is more tricky since we need to switch the limit with the $\inf$. We write this as:
\begin{align}
\limsup_{\epsilon \rightarrow 0} \inf_{u' \in \mathcal{V}_\psi(\mathcal{A}')} \inf_{\chi \in \mathscr{H}_P} - \frac{1}{\epsilon}
\ln \frac{ | \left< \chi \right| e^{-a\sin(2\pi \epsilon) P} u' \left| \psi \right> |}{ \| \chi \|}
\end{align}
where we are free to restrict to the dense subspace $\mathscr{H}_P = \cup_{\Lambda > 0} \Pi_\Lambda(P) \mathscr{H}$ of vectors that are entire analytic for $e^{ i z P}$ while
still achieving the norm.
We can then shift $\chi \rightarrow e^{a\sin(2\pi \epsilon) P} \chi$ then use scale invariance to restrict to normalized $\chi$ and apply Jensen's inequality
$e^{ \alpha \left< \chi \right| P \left| \chi \right>} \leq \| e^{ \alpha P} \chi \| \leq e^{ \alpha \Lambda }$ so that:
\begin{equation}
\geq \sup_{r > 2} \inf_{\chi \in (\mathscr{H}_P)_{= 1}}  \left( - r \ln {F(\chi,\psi;\mathcal{A})} + {a r \sin(\frac{2\pi}{r})} {\left< \chi \right| P \left| \chi \right> } \right)
\end{equation}
after which we can relax to $\chi \in \mathscr{H}_{\leq 1}$ at the cost of an extra $\geq$. 
We have also redefined $r = 1/\epsilon$ and used monotonicity in $r$ to convert the limit to a supremum.
Now we can write these in terms of the state in the natural cone $\xi_\chi$ using Lemma 5(b) of \cite{Faulkner:QNECwithoutAssumptions} for the expansion $\Theta(\chi|_{\mathcal{A}})$
\begin{equation}
\| \sqrt{P} \chi \|^2 \geq \frac{1}{2}\| \sqrt{P} \xi_\chi \|^2\,, \quad - \ln F(\chi,\psi;\mathcal{A}) \geq \frac{1}{2}  \left( 1 - {\left< \xi_\chi \right| \left. \xi_\psi \right>} \right)^2
\end{equation}
where $\xi_\chi \in \mathscr{P}^\natural(\mathcal{A})$. Leaving:
\begin{equation}
\geq \sup_{r > 2} \inf_{\chi \in \mathscr{P}^\natural(\mathcal{A}')_{\leq 1}}  \left(  \frac{r}{2}  \left(1 - \left< \chi \right| \left. \xi_\psi \right> \right)^2+ \frac{a r}{2}\sin(\frac{2\pi}{r}) \left< \chi \right| P \left| \chi \right> \right)
\end{equation}
where we have relaxed to $\| \chi \| \leq 1$ at the cost of another $\geq$. 
At this point we have a function
$(r,\chi)\rightarrow f(r,\chi)\in\mathbb{R}_{\geq 0}\cup\{\infty\}$
with $r>2$ and $\chi\in \mathscr{K}\equiv
\mathscr{P}^\natural(\mathcal{A}')_{\leq 1}$.
The set $\mathscr{K}$ is weakly compact since the unit ball is weakly compact
and the natural positive cone is closed and convex.
The function $f(r,\chi)$ is weakly lower semicontinuous in $\chi$,
since the positive quadratic form
$\langle\chi|P|\chi\rangle$ is weakly lower semicontinuous.
It is also non-decreasing in $r$, since both $r$ and
$r\sin(2\pi/r)$ are increasing for $r>2$.

These properties suffice to exchange the supremum and infimum.
Indeed, write $m=\sup_{r>2}\inf_{\chi\in \mathscr{K}}f(r,\chi)$.
If $m<\infty$, then for any $c>m$ the sets
\begin{equation}
\mathscr{K}_r(c)=\{\chi\in \mathscr{K}:f(r,\chi)\leq c\}
\end{equation}
are nonempty, weakly closed and compact, and nested as $r$ increases.
Cantor's intersection theorem gives a nonempty intersection, so
$\inf_{\chi\in \mathscr{K}}\sup_{r>2}f(r,\chi)\leq c$.
Taking $c\rightarrow m^+$ gives
\begin{equation}
\inf_{\chi\in \mathscr{K}}\sup_{r>2}f(r,\chi) \leq \sup_{r>2}\inf_{\chi\in \mathscr{K}}f(r,\chi)\leq \inf_{\chi\in \mathscr{K}}\sup_{r>2}f(r,\chi)
\end{equation}
implying equality. 
The case $m=\infty$ follows immediately from the same general
inequality.
The $\sup_r$ is now easy
to compute since it is guaranteed to give $=\infty$ unless $\left< \chi \right| \left. \xi_\psi \right> = 1$
which, for $\| \chi \| \leq 1$, is only possible when $\chi = \xi_\psi$ so we complete
the bound $\geq 2\pi a\,\Theta(\psi|_\mathcal{A})$ and hence, combined with the upper bound above, equality:
\begin{equation}
\lim_{p' \rightarrow 2^+} H_{p'}(\psi|_{ \mathcal{A}|\mathcal{A}_a}) = H_2(\psi|_{ \mathcal{A}|\mathcal{A}_a})
\end{equation}
\end{proof}

\subsection{Dynamical R\'{e}nyi Entropy}
\label{sec:dynamical}

At $p=2$ this conditional entropy has an interpretation as the dynamical conditional entropy in a semiclassical approximation of gravity with quantum fields
cut along a Killing horizon at zeroth order \cite{Hollands:2024dynamical}.
The appropriate dynamical entropy in this limit is:
\begin{equation}
\label{eq:Sdyn}
S^{\rm dyn}(V:V_h) = - D(\psi|_V) + 2\pi (V_h- V) \Theta(\psi|_V)
\end{equation}
 where $V_h$ is the value of the affine parameter at the Killing horizon. Usually
the (teleologically defined) black hole entropy is the area of the event horizon divided by $4G_N$, plus quantum corrections from entanglement across the horizon which
in turn is approximately
$ - D(\psi|_V)$ up to some constant additive shift. The extra term in the dynamical entropy comes from picking the apparent horizon (chosen using infalling null curves from the past asymptotic boundary) to track the generalized entropy. There are several motivations for this choice, including a coarse-graining procedure discussed in \cite{Engelhardt:2017aux,Engelhardt:2018kcs}.
Then:
\begin{equation}
H_2(\psi|_{\mathcal{A}_V|\mathcal{A}_{V_h}}) =  S^{\rm dyn}(V : V_h) - S^{\rm dyn}(V_h : V_h)
\end{equation}
where $V_h > V$. Note that the sign constraint $\geq 0$, which is a little strange for a conditional entropy\footnote{For example the renormalized conditional entropy in the presence of a conditional
expectation is $- D(\rho | \rho \circ E) \leq 0$ \cite{Gao:2019subalgebras}.} comes from the funny sign of $V_h > V$ - that is we are considering
the state of quantum fields to the past of the Killing horizon including the white hole region.
We might define a dual conditional entropy that works now for $V > V_h$, so that now $\mathcal{A}_V$ is to the future of the Killing horizon\footnote{A useful way to remember our notation is that since $V_h$ will be associated to the Killing horizon there are no shifts
from the quantum expansion terms since the Killing horizon is locally stationary in the area. Hence $\Theta$ terms will only arise at $V$.}, using:
\begin{align}
\widetilde{H}_2(\psi|_{\mathcal{A}_{V_h}|\mathcal{A}_{V}})
&\equiv -  H_2(\psi|_{\mathcal{A}_{V}'|\mathcal{A}_{V_h}'})  =  S^{\rm dyn}(V_h : V_h)-S^{\rm dyn}(V: V_h) \leq 0 
\end{align}
(where in order to derive the second equality we assumed $D(\psi|_{V_h}) < \infty, D(\psi|_{V'}) < \infty$
so we could apply the sum rule \eqref{eq:sumrule}.)
Now the conditional entropy is negative and represents the increase of the dynamical entropy, that is the second law.  
Hence this dual quantity has a more straightforward physical interpretation. 

Given this discussion it is natural to generalize these quantities away from $p=2$.
Let us start by giving a natural generalization of the dual conditional dynamical entropy
for $V> V_h$ and $\psi \in L_p(\mathcal{A}_{V_h}')$:
\begin{align}
\widetilde{H}_p(\psi|_{\mathcal{A}_{V_h}|\mathcal{A}_{V}})
&\equiv -  H_p( \Delta_{\eta;\mathcal{A}_{V_h}}^{1/p-1/2}( \psi) |_{\mathcal{A}_{V}'|\mathcal{A}_{V_h}'})  
\end{align}
where the re-weighting achieves unitary independence under $\psi \rightarrow  u'\psi$ for $u' \in \mathcal{U}( \mathcal{A}_{V_h}')$
so that the notation:
\begin{equation}
\widetilde{H}_p(\rho|_{\mathcal{A}_{V_h}|\mathcal{A}_{V}})
\equiv \widetilde{H}_p(\xi_\rho|_{\mathcal{A}_{V_h}|\mathcal{A}_{V}})
\end{equation}
is still appropriate where $\rho \in (\mathcal{A}_{V_h})^+_\star$. 
We also consider the dynamical R\'{e}nyi entropy still for $V > V_h$:
\begin{align}
\label{eq:Spdyn}
S_p^{\rm dyn}(V : V_h)&  \equiv - \frac{2p}{p-2} \ln \| \Delta_{\eta;\mathcal{A}_{V_h}}^{1/p-1/2}(\psi) \|_{L_p(\mathcal{A}_V \subset \mathcal{A}_{V_h})}
\end{align}
This is essentially the original norm without the action of $J^p$. While this seems innocuous enough, various putative properties
become more difficult to prove so we leave this for future work. We do comment that taking the limit to $p \rightarrow 2$ naively, in particular without attempting to rigorously justify the answer, one finds $\lim_{p \rightarrow 2} S_p^{\rm dyn}(V:V_h) = S^{\rm dyn}(V: V_h) $.
Also it is clear that $S_p^{\rm dyn}(V_h:V_h) = - D_p(\psi|_{V_h})$ and that $V \rightarrow  S_p^{\rm dyn}(V:V_h)$ is non-decreasing for $V > V_h$.
These are all desirable properties for the R\'{e}nyi generalization of the dynamical entropy.

\section{\texorpdfstring{$L_p$}{Lp} Quantum Expansions}
\label{sec:expansions}

In terms of the $L_p$ conditional quantum expansions defined in the introduction we note that
we can write:
\begin{equation}
\label{eq:Thetapa}
H_p(\psi|_{\mathcal{A}|\mathcal{A}_a}) = 2\pi a\, \Theta_p( \psi|_{\mathcal{A}} ; a)
- D_p(\psi|_{\mathcal{A}}) +D_p(\psi|_{\mathcal{A}_a})
\end{equation}
where this defines $ \Theta_p( \psi|_{\mathcal{A}} ; a)$ for $a > 0$.
This quantity allows us to constrain the limit to $a \rightarrow 0$. We also recall here the definition of the non-conditional $L_p$ expansion
defined for $ 2 \leq p < \infty$
\begin{equation}
\label{eq:Thetap}
\Theta_p(\psi|_{\mathcal{A}}) \equiv \frac{p \sin(2\pi/p)}{\pi(p-2)} \inf_{ u' \in \mathcal{U}(\mathcal{A}')} \frac{\| \sqrt{P} u' \varphi \|^2}{ \| \varphi \|^2}
= \frac{p \sin(2\pi/p)}{2\pi(p-2)} \frac{\| \sqrt{P} \varphi \|^2}{ \| \varphi \|^2}
\end{equation}
where $\varphi$ is uniquely defined by the polar decomposition of $\psi = \theta' \Delta_{\eta,\varphi;\mathcal{A}}^{-1/p} \eta$. At $p=2$ the prefactors are understood by continuity. 

\begin{remark}
\label{rem:pinfty}
At $p = \infty$ we set $\Theta_\infty(\psi|_{\mathcal{A}}) \equiv 0$. This is the only consistent choice, since at $p=\infty$ everything is trivial along the HSMI.
If $\psi \in L_\infty(\mathcal{A}_{V_0}')$ then $\psi = a' \eta$ for some $a' \in \mathcal{A}_{V_0}' \subset \mathcal{A}_V'$, $V \geq V_0$, and this $a'$ is unique since $\eta$ is separating for $\mathcal{A}_V'$.
Hence $\| \psi \|_{L_\infty(\mathcal{A}_V')} = \| a' \|$ and $D_\infty(\psi|_V) = 2 \ln \| a' \|$ is independent of $V \geq V_0$. 
Similarly, at $p=\infty$ we have $\sin(2\pi/p)=0$, so the translation in \eqref{eq:HSMIvar} is unitary and $\Theta_\infty(\psi|_V;a) = 0$ for all $a>0$.
\end{remark}

The following result
follows from the log-convexity of the conditional norms in $a$:
\begin{lemma}
  \label{lem:log-convexity}
For $2 < p \leq \infty$, consider some $\psi \in L_p(\mathcal{A}')$. Then 
$\Theta_p(\psi|_{ \mathcal{A}} ; a) \in  \mathbb{R}_{\geq 0}$ is a monotonically non-increasing function of $a$
with the bounds:
\begin{equation}
 \frac{D_p(\psi|_\mathcal{A}) - D_p(\psi|_{\mathcal{A}_a})}{2\pi a}  \leq \Theta_p(\psi|_{ \mathcal{A}} ; a)  \leq  \Theta_p(\psi|_\mathcal{A})
\end{equation}
\end{lemma}
\begin{proof} 
We first note that log-convexity of the conditional norm in $a$, (2) of Lemma~\ref{lem:HSMInorm}, implies
that $\Theta_p(\psi|_{\mathcal{A}} ; a)$ is a monotonically decreasing function of $a$.
Hence the following limit exists but could be $\infty$:
\begin{equation}
\lim_{ a\rightarrow 0^+} \Theta_p(\psi|_{\mathcal{A}} ; a) = \sup_{a > 0} \Theta_p(\psi|_{\mathcal{A}} ; a)
\geq \Theta_p(\psi|_{\mathcal{A}} ; a)  \geq  \frac{D_p(\psi|_\mathcal{A}) - D_p(\psi|_{\mathcal{A}_a})}{2\pi a} %
\end{equation}
where we used positivity, (3) of Lemma~\ref{lem:Hpbasic}.
For the upper bound, we first replace $\psi$ by $(\theta')^\dagger\psi$ using (3) of Lemma~\ref{lem:condbasic}.
The ordinary and conditional norms are unchanged, as is $\varphi$, so both quantum expansions are unchanged and we can assume
$\psi = \Delta_{\eta,\varphi;\mathcal{A}}^{-1/p}\eta$.
We can then exchange the $\inf$ and $\sup$ in the definition of $\Theta_p(\psi|_{\mathcal{A}} ; a)$ to get:
\begin{equation}
\lim_{ a\rightarrow 0^+} \Theta_p(\psi|_{\mathcal{A}} ; a) \leq \inf_{\chi \in \mathscr{H}_1, u' \in \mathcal{U}(\mathcal{A}')}\sup_{a > 0}-  \frac{p}{\pi a(p-2)} \ln\frac{ \| 
U( a e^{-2\pi i/p} )u' \Delta_{\eta,\chi;\mathcal{A}}^{1/p-1/2} \psi \|}{ \| \psi \|_{L_p(\mathcal{A}')}}
\end{equation}
Choose $\chi = \varphi/\|\varphi\|$, which is an optimizer by Lemma~\ref{lem:normoptimizer}, so that
$\Delta_{\eta,\chi;\mathcal{A}}^{1/p-1/2}\psi/\|\psi\|_{L_p(\mathcal{A}')} = \varphi/\|\varphi\|$.
Bounding the infimum by this choice, we can compute $\sup_{a>0}$ and then take the $\inf_{u'}$ using \eqref{eq:Thetap}, giving the claim.
\end{proof}

We now consider the dependence on $p$. 
\begin{lemma}
\label{lem:Thetap-p}
Consider some $\psi \in L_p(\mathcal{A}')$ with $\| \psi \| = 1$ for $2 < p \leq \infty$. Then
$p' \rightarrow \Theta_{p'}(\psi|_{\mathcal{A}};a)$ is continuous for $2 \leq p' \leq p$
with the limit at $p' = 2$ equal to the standard quantum expansion:
\begin{equation}
\lim_{p' \rightarrow 2} \Theta_{p'}(\psi|_{\mathcal{A}};a) = \Theta(\psi|_{\mathcal{A}})
\end{equation}
Note that when $\psi \in L_\infty(\mathcal{A}')$ it is the case that $\Theta_{\infty}(\psi|_{\mathcal{A}};a) = 0$. 
\end{lemma}
\begin{proof}
Continuity follows from continuity in $p'$ of $H_{p'}(\psi|_{\mathcal{A}|\mathcal{A}_a})$ and $D_{p'}$, see (5) of Lemma~\ref{lem:Hpbasic}; the limit follows from (5) of Lemma~\ref{lem:HpHSMI} and the definition \eqref{eq:H2def}.
\end{proof}

We have a necessary and sufficient condition for the quantum expansion to vanish:

\begin{lemma} 
\label{lem:vanishing}
Fix some $2 \leq p < \infty$ with $\psi \in L_p(\mathcal{A}')$ and $\| \psi \| = 1$.
Assume $\eta$ is the unique $P$ vacuum. Then $\Theta_p(\psi|_{\mathcal{A}}; a) = 0$ iff $\omega_\psi|_{\mathcal{A}} = \omega_\eta|_{\mathcal{A}}$.
Also $\Theta_p(\psi|_{\mathcal{A}}) = 0$ iff $\omega_\psi|_{\mathcal{A}} = \omega_\eta|_{\mathcal{A}}$. 
\end{lemma}
\begin{proof}
  The reverse implication is clear. 
$\Theta_p(\psi|_{\mathcal{A}}; a) = 0$ implies that $D_p(\psi|_{\mathcal{A}_a}) = D_p(\psi|_\mathcal{A})$. We can 
then use the integral formula $0 = 2\pi \int_0^a d b\, \Theta_p(\psi|_{\mathcal{A}_b})$ (this is \eqref{mainintegral}, proven in Section~\ref{sec:proofs}; its proof does not depend
on this Lemma). But then $\Theta_p(\psi|_{\mathcal{A}_b}) = 0$ almost everywhere for $0 < b < a$ and hence everywhere on this interval by convexity. Taking the limit $b \rightarrow 0^+$, using lower semicontinuity from the right, gives $\Theta_p(\psi|_\mathcal{A})=0$ (see Lemma~\ref{lem:Thetacont}).
So $\sqrt{P} \varphi = 0$ implying $\varphi = \eta$ and hence the result. 

If $p=2$ we simply have $\sqrt{P} \varphi =0 \implies \varphi = \eta$ and the result.
Similarly for $\Theta_p(\psi|_{\mathcal{A}}) =0 \iff \varphi = \eta$ . 
\end{proof}

We will need some continuity statements for varying $\psi$:
\begin{lemma}
\label{lem:Thetacont}
Suppose that $\psi_i \rightarrow \psi$ in $L_p(\mathcal{A}')$. Then:
\begin{enumerate}
\item For $p>2$, $\lim_i \Theta_p(\psi_i|_{\mathcal{A}};a) = \Theta_p(\psi|_{\mathcal{A}};a)$
\item $\liminf_i \Theta_p(\psi_i|_{\mathcal{A}}) \geq \Theta_p(\psi|_{\mathcal{A}})$.
\end{enumerate}
Thus $\liminf_{a \rightarrow 0^+} \Theta_p(\psi|_{\mathcal{A}_a}) \geq \Theta_p(\psi|_{\mathcal{A}})$, giving lower semicontinuity from the right. 
\end{lemma}
\begin{proof}
(1) Continuity is clear from continuity of the $L_p$ conditional norm
which follows from the reverse triangle inequality and the bound $\| \psi \|_{L_p(\mathcal{A}' \subset \mathcal{A}_a')}
\leq \| \psi \|_{L_p(\mathcal{A}_a')} \leq  \| \psi \|_{L_p(\mathcal{A}')}$.
(2) We have continuity of $\varphi_i \rightarrow \varphi$ in $\mathscr{H}$ from Lemma~\ref{lem:polarcont}
so lower semicontinuity follows from estimating the expectation in a spectral cut of $P$: $P \Pi_\Lambda \leq P$.
The last result applies (2) to $\psi_i = U(a_i) \psi$. 
\end{proof}

\section{Proof of the Main Theorems}
\label{sec:proofs}

We aim to show the inequalities in Lemma~\ref{lem:log-convexity} are in fact saturated as $a \rightarrow 0$.
 In order to do this we need a new method to compute the $V$ derivative of $D_p(\psi|_V)$.
 We will also need to show that $D_p(\psi|_V)$ is appropriately differentiable and continuous. We aim for:
\begin{lemma}
  \label{lem:finalderiv}
Fix $2 < p < \infty$. For all $\psi \in L_p(\mathcal{A}_{V_0}')$, $V \rightarrow D_p(\psi|_V)$ is differentiable and locally Lipschitz 
for $V > V_0$ with:
\begin{equation}
- \partial_V D_p(\psi|_{V}) = 2\pi \lim_{a \rightarrow 0^+} \Theta_p(\psi|_{V};a) = 2\pi \Theta_p(\psi |_V)
\end{equation}
Thus $D_p(\psi|_{V})$ satisfies left endpoint convexity for $V > V_0$ and $a > 0$:
\begin{equation}
- \partial_V D_p(\psi|_{V}) \geq \frac{D_p(\psi|_V) - D_p(\psi|_{V+a})}{a} 
\end{equation}
\end{lemma}

We need some preliminary results. Firstly we will need to allow for $L_p$ spaces with $1 < p < \infty$.
For $1 < p < 2$ we briefly summarize some results we will need following the Araki-Masuda formalism. It will be sufficient to define these
as Banach dual spaces, $L_{p}(\mathcal{A}') \equiv L_{p'}(\mathcal{A}')^\star$ where $1/p+1/p'=1$. These spaces
\cite{araki1982positive} are reflexive (for $p\neq 1$) so the double dual gives back the original $L_{p'}$ space.
The natural sesquilinear pairing $\left< \cdot, \cdot \right> : L_{p}(\mathcal{A}') \times L_{p'}(\mathcal{A}') \rightarrow \mathbb{C}$ can be used to compute the norm:
\begin{equation}
\| \psi \|_{L_p(\mathcal{A}')} = \sup_{\zeta \in L_{p'}(\mathcal{A}')_{\leq 1} } |\left< \zeta, \psi \right>|
\end{equation}
where the new positive cones $\psi \in L_p^+(\mathcal{A}')$ satisfy $\left< \zeta, \psi \right> \geq 0$ for all
$\zeta \in L_{p'}^+(\mathcal{A}')$. 
These spaces can be understood as extensions of $\mathscr{H}$ and so general elements cannot be written using 
$\mathscr{H}$ as we did for $2 \leq p \leq \infty$. Nevertheless,
the multi-variable KMS functionals worked out in \cite{araki1982positive} for the relative modular operator allow us to justify formal
expressions such as the polar decomposition $\theta' \Delta_{\eta,\varphi}^{-1/p} \eta \in L_p(\mathcal{A}')$ via:\footnote{Recall that the domain condition $\eta \in \mathcal{D}(\Delta_{\eta,\varphi}^{-1/2})$ does not generally extend beyond $-1/2$ so these cannot always be understood as elements in $\mathscr{H}$.
}
\begin{equation}
\left< \theta' \Delta_{\eta,\varphi}^{-1/p} \eta, \phi' \Delta_{\eta,\chi}^{-1/p'} \eta \right>
= \left< \eta \right| \Delta_{\eta, \varphi}^{-1/p} (\theta')^\dagger \phi' \Delta_{\eta,\chi}^{-1/p'} \left| \eta \right> 
\end{equation}
This also allows us to extend the H\"{o}lder product $L_{q}(\mathcal{A}') \times_{\mathcal{A}'} L_r(\mathcal{A}')$ to $0 \leq 1/q+1/r \leq 1$ with corresponding formal expressions. The R\'{e}nyi divergence is defined in the same way as for $2 \leq p < \infty$,
$D_p(\psi|_{\mathcal{A}}) = 2p/(p-2) \ln \| \psi \|_{L_p(\mathcal{A}')}$, and still satisfies $D_p(\psi|_{\mathcal{A}}) \geq 0$ for normalized states. 

The contractive subspace $L_{p'}(\mathcal{A}') \subset L_{p'}(\mathcal{B}')$ for $\mathcal{A}' \subset \mathcal{B}'$ 
and $2 \leq p' \leq \infty$ does not extend to the dual spaces $1 < p < 2$. Instead we will only need to work with the contractions from Lemma~\ref{lem:Tz}. In the HSMI case they can be turned into contractions on the same $L_{p'}$ space, $U(\alpha) = e^{-i \alpha P} : L_{p'}(\mathcal{A}' )\rightarrow L_{p'}(\mathcal{A}')$ for $2 \leq p' \leq \infty$ where $\alpha = V + e^{-2\pi i/p'} V_{p'}$ with $V,V_{p'} \geq 0$, see \eqref{eq:wedge}. 
These contractions can then be passed 
to $1 < p < 2$ using the transpose contraction. We will also need the action of $P : L_{p'}(\mathcal{A}' )\rightarrow L_{p'}(\mathcal{A}')$ as an unbounded Banach space operator, and its transpose: 
\begin{lemma}
\label{lem:transpose}
For $1 < p \leq 2$ the transpose map $U(-\alpha^\star)^\star : L_{p'}(\mathcal{A}')^\star
\rightarrow L_{p'}(\mathcal{A}')^\star$, with $L_{p'}(\mathcal{A}')^\star = L_p(\mathcal{A}')$, which we will formally write as $U(\alpha) = e^{-i \alpha P}$, is a contraction-valued strongly holomorphic function inside the wedge:
\begin{equation}
\| U(\alpha) \psi \|_{L_p(\mathcal{A}')} \leq \| \psi \|_{L_p(\mathcal{A}')}\, \quad \alpha = - V - e^{-2\pi i/p} V_p\, \quad V, V_p \geq 0
\end{equation}
For $V = V_p$, the positive cones are preserved. The map takes the form
\[
U(\alpha) = \exp( - i \epsilon P e^{-\pi i /p} ), \qquad \epsilon >0,
\]
for all $1 < p < \infty$. 

Additionally $P : L_{p'}(\mathcal{A}') \rightarrow L_{p'}(\mathcal{A}')$ is a closable densely defined operator for $2 \leq p' < \infty$ where the transpose $P^\star : L_{p}(\mathcal{A}') \rightarrow L_{p}(\mathcal{A}')$ is densely defined, commutes with $U(\alpha)$ and agrees with $P$ when restricted to $\mathcal{D}(P)\cap L_p(\mathcal{A}')$. For notational convenience we will conflate $P^\star$ with $P$.
\end{lemma}
\begin{proof}
The transpose map of a contraction is a contraction. $\alpha \rightarrow U(\alpha) \psi \in L_p(\mathcal{A}')$ is a Banach-space-valued analytic function
for $\alpha$ in this wedge-like region. 
This follows from the fact that weak analyticity implies strong analyticity.
The positive cone condition 
passes straightforwardly to the transpose. 

A dense domain for $P$ is established with
 $U(\epsilon e^{-i \pi/p}) L_p(\mathcal{A}')$ with $\epsilon \rightarrow 0$
 for all $1 < p < \infty$.
\end{proof}

Given $\psi \in L_p(\mathcal{A}')$ we can form $\psi^\dagger \psi \equiv J^p_{\mathcal{A}'}(\psi) \times_{\mathcal{A}'} \psi \in L_{p/2}^+(\mathcal{A}')$ which removes the unimportant unitary in the polar decomposition $\theta'$. 
The $L_{p/2}$ norm of $\psi^\dagger \psi$ is then the square of the $L_p$ norm of $\psi$. The $L_p$ norm
for the subalgebra can then be computed with the positive contraction. Using Lemma~\ref{lem:productincompat} with $\zeta = \psi$ (in the extended form of Remark~\ref{rem:therelation}, since $2/p$ can exceed $1/2$) and $T_{1/p}(\mathcal{A}' \subset \mathcal{A}_a') = U(a(e^{-2\pi i/p}-1))$ we have
$J^p_{\mathcal{A}_a'}(\psi) \times_{\mathcal{A}_a'} \psi = \exp(i a (1-e^{-i 2\pi /p}) P)J^p_{\mathcal{A}'}(\psi) \times_{\mathcal{A}'} \psi$.
So that:
\begin{equation}
\label{eq:normshift}
\| \psi \|^2_{L_{p}(\mathcal{A}_a')} = \| J^p_{\mathcal{A}_a'}(\psi) \times_{\mathcal{A}_a'} \psi \|_{L_{p/2}(\mathcal{A}_a')}  =  \left\| \exp(-i a e^{-i 2\pi /p} P) \psi^\dagger \psi \right\|_{L_{p/2}(\mathcal{A}')}
\end{equation}
Note that $ \exp(-i a e^{-i 2\pi /p} P)   : L_{p/2}^+(\mathcal{A}') \rightarrow L_{p/2}^+(\mathcal{A}')$
preserves the positive cone. 
This map was important for formulating the QNEC for integer $p/2$ in \cite{Kibe:2026wsg}.

We also need the following result on perturbation theory for the $L_p$ norms taken from Kosaki \cite{Kosaki:1984uniform}:
\begin{lemma}
\label{lem:kosaki}
Fix some $1 < p < \infty$ and consider a one-parameter family $\varrho(t) \in  L_p^+(\mathcal{A}')$ for $t \geq 0$ with some $ \varrho'_0\in L_p(\mathcal{A}')$ such that
\begin{equation}
\lim_{t \rightarrow 0^+} \| (\varrho(t) - \varrho(0))/t - \varrho'_0 \|_{L_p(\mathcal{A}')} = 0
\end{equation}
Then:
\begin{equation}
\lim_{ t \rightarrow 0^+} \frac{ \| \varrho(t) \|^p_{L_p(\mathcal{A}')} - \| \varrho(0) \|^p_{L_p(\mathcal{A}')}}{t}
= p {\rm Re} \left< \varphi_0 \right| \Delta_{\eta,\varphi_0;\mathcal{A}}^{-1/2+1/p} \left|\varrho'_0\right>
\end{equation}
where $\varrho(0) = \Delta_{\eta,\varphi_0;\mathcal{A}}^{-1/p} \eta$ is the polar decomposition of the endpoint.
\end{lemma}
\begin{proof}
See \cite{Kosaki:1984uniform} and the duality and uniform convexity results in \cite[Sections~8-10]{araki1982positive}. The Haagerup/Kosaki form of the weighted $L_p$ spaces was used here. But these are known to be equivalent \cite{araki1982positive,kosaki1984applications}.
\end{proof}

We are now ready to prove the derivative result: 
\begin{proof}[Proof of Lemma~\ref{lem:finalderiv}]
  Without loss of generality set $V_0=0$. 
Note that $U(\alpha) \psi^\dagger \psi$ has an analytic continuation as an $L_{p/2}(\mathcal{A}')$ valued function, inside the wedge in a neighborhood of the point 
$ \alpha = V e^{ -2\pi i/p}$.
Hence we can apply the Cauchy integral formula to show the existence of:
\begin{equation}
PU(V e^{-2\pi i/p}) \psi^\dagger \psi = \oint_{C_{\delta}(V e^{-2\pi i/p})} \frac{d \alpha}{2\pi (\alpha-V e^{-2\pi i/p})^2}  U(\alpha) \psi^\dagger \psi \in L_{p/2}(\mathcal{A}')
\end{equation}
where this is well defined by \eqref{eq:wedge} (for $p \geq 4$) and Lemma~\ref{lem:transpose} (for $p<4$), since $V e^{-2\pi i/p}$ lies in the interior of the region of contractions on $L_{p/2}(\mathcal{A}')$ (for $p=4$ this region is the lower half-plane ${\rm Im}\, \alpha \leq 0$).

The Cauchy integral formula of course computes the derivative meaning:
\begin{equation}
\lim_{ h \rightarrow 0} \| (U((V+h)e^{-2\pi i/p}) - U(V e^{-2\pi i/p}))/h \psi^\dagger \psi + i e^{-2\pi i/p} PU(V e^{-2\pi i/p}) \psi^\dagger \psi \|_{L_{p/2}(\mathcal{A}')} = 0
\end{equation}
in both directions for $h>0, h<0$. In particular we estimated the above difference using the Cauchy integral and bounded the difference
using the triangle inequality on the norm. We can then use Lemma~\ref{lem:kosaki} (for $L_{p/2}$, applied for both signs of $h$) together with \eqref{eq:normshift} to compute the derivative for $V > 0$:
\begin{align}
- \partial_{V} D_p(\psi|_{V})
&=  \frac{p}{2-p} {\rm Im}\left( e^{-2\pi i/p} \left< \varphi^{(V)} \right| \Delta_{\eta,\varphi^{(V)};\mathcal{A}}^{2/p-1/2} PU(V e^{-2\pi i/p}) \psi^\dagger \left. \psi \right>\right)/
\|\psi \|^p_{L_{p}(\mathcal{A}_V')}
\end{align}
where the polar decomposition comes from $\Delta_{\eta,\varphi^{(V)};\mathcal{A}}^{-2/p} \eta = U(V e^{-2\pi i/p})\psi^\dagger \psi$. Here $\varphi^{(V)} = U(V)\varphi_V \in \mathscr{P}^\natural(\mathcal{A})$, with $\varphi_V$ as in the introduction, and the two vectors have the same norm and $P$ expectation. Hence $D_p(\psi|_V)$ is differentiable. We can also estimate the derivative as:
\begin{align}
- (1/2-1/p) \partial_{V} D_p(\psi|_{V}) &\leq \frac{1}{2} \| P U(V e^{-2\pi i/p}) \psi^\dagger \psi \|_{L_{p/2}(\mathcal{A}')}
/\| \psi \|^2_{L_p(\mathcal{A}_V')} \\ & \leq \frac{1}{2} \| P U(\epsilon e^{-2\pi i/p})\psi^\dagger \psi \|_{L_{p/2}(\mathcal{A}')}/\|\psi \|^2 
\end{align} 
where $\epsilon >0$. We used the fact that $P$ and $U(\alpha)$ commute, and the contractive property of the latter. We also used monotonicity of the $L_p$ norms with respect to $p$ to estimate the denominator. This gives a Lipschitz function for $V\geq \epsilon$ and hence a locally Lipschitz function.

Furthermore the derivative can be computed as follows:
\begin{align}
\label{eq:rhs}
- \partial_{V} D_p(\psi|_{V}) &=  \frac{p}{2-p} \frac{{\rm Im} \left( e^{-2\pi i/p}  \left< \varphi^{(V)} \right| \Delta_{\eta,\varphi^{(V)};\mathcal{A}}^{2/p-1/2}  P  \Delta_{\eta,\varphi^{(V)};\mathcal{A}}^{-2/p+1/2} \left| \varphi^{(V)}\right> \right)}{ \| \varphi^{(V)} \|^2 } \\
& = \frac{p}{2-p} {\rm Im} \left(e^{-2\pi i/p} \left< \zeta \right| \Delta_{\eta,\zeta;\mathcal{A}}^{is} P \Delta_{\eta,\zeta;\mathcal{A}}^{-is} \left| \zeta \right>\right) |_{is = 2/p-1/2}
\end{align}
where $\zeta = \varphi^{(V)} / \| \varphi^{(V)}\|$.

It is known \cite{Ceyhan:2018zfg} how to compute the $s$ dependence of these expectations. These formulas were
derived under the assumption that $\zeta \in \mathcal{D}(\sqrt{P})$. In fact all we need to know is that
there is a single value of $s$ where this expectation is finite.  We give the updated argument here:

Set $\zeta_s = \Delta_{\eta,\zeta}^{-is} \zeta $. 
Consider $\gamma_\epsilon(z,s) = \left< \zeta_s \right| (1-e^{ i\epsilon e^{2\pi (z+s)} P})/\epsilon \left| \zeta_s \right>$
initially for real $s$ and $0 \leq {\rm Im} z \leq 1/2$. 
Monotone convergence establishes the existence of the limit $e^{2\pi h} P_s = \lim_{\epsilon \rightarrow 0 } \gamma_\epsilon(i/4+h,s) = e^{2\pi (s+h)} \left< \zeta_s \right| P \left| \zeta_s \right> \in [0,\infty]$ for real $s, h$. In \cite{Ceyhan:2018zfg,Faulkner:QNECwithoutAssumptions} it was shown that $\gamma_\epsilon(z,s)$ has an analytic continuation to the tube based on the rhombus shaped region
 $-{\rm Im} z <{\rm Im}s < 1/2 - {\rm Im} z, 0 < {\rm Im} z < 1/2$ where $|1-\epsilon \gamma_\epsilon(z,s)| \leq 1$. 

Now since we know that $\Delta_{\eta,\zeta}^{-is} \zeta \in L_{q_s}(\mathcal{A}')$ where $1/q_s = 1/2-{\rm Im } s$
and $\Delta_{\eta,\zeta}^{-is^\star} \zeta \in L_{q_s'}(\mathcal{A}')$ where $1/q_s'+1/q_s=1$, 
we can give an alternative proof of this analytic continuation using the $L_{q_s}$ dual pairing:
\begin{equation}
\gamma_\epsilon(z,s) = \frac{1}{\epsilon}\left<\Delta_{\eta,\zeta}^{-is^\star} \zeta, \left(1- U(- \epsilon e^{2\pi (z+s)}) \right) \Delta_{\eta,\zeta}^{-is} \zeta  \right> 
\end{equation}
where $- \epsilon e^{2\pi (z+s)} = V + V_{q_s} e^{- 2\pi i/ q_s}$ for $V,V_{q_s} \geq 0$ in the specified range
with ${\rm Im}\, s \geq 0$ (for ${\rm Im}\, s < 0$ we have $q_s < 2$ and instead $- \epsilon e^{2\pi (z+s)} = -V - V_{q_s} e^{- 2\pi i/ q_s}$ as in Lemma~\ref{lem:transpose}), 
so that $U(- \epsilon e^{2\pi (z+s)}) : L_{q_s}(\mathcal{A}') \rightarrow L_{q_s}(\mathcal{A}')$ is a contraction. 
We have established above, via \eqref{eq:rhs}, that the limit exists and is finite at $s = i(1/2-2/p)$ where $p$ is fixed in the range $2 < p < \infty$, 
which lies inside the rhombus (for $2<p<4$ this point has ${\rm Im}\, s < 0$ and $q_s = p/2 < 2$).
Since (a) $e^{ - \gamma_\epsilon(z,s)}$ is uniformly bounded and non-vanishing inside the rhombus, (b) the limit exists at ${\Im s} =0$ and $z = i/4+ h$, and (c) the limit exists and is non-zero at one point, the limit must exist and be non-zero at all points inside the rhombus via an application of the Montel/Vitali-Porter theorem (with the identity set
equal to: ${\rm Im} s = 0, {\rm Im z}= 1/4$) and Hurwitz's theorem. It follows that $P_s < \infty$ and the explicit form follows from (3.34) of \cite{Faulkner:QNECwithoutAssumptions} for real $s$ (using the fact that $\zeta \in \mathcal{D}(\sqrt{P})$), extended into the rhombus by analyticity:
\begin{equation}
 2 \| \varphi^{(V)} \|^2 \left< \zeta \right| \Delta_{\eta,\zeta;\mathcal{A}}^{is} P \Delta_{\eta,\zeta;\mathcal{A}}^{-is} \left| \zeta \right>
 = (e^{-2\pi s} +1)P_{\varphi^{(V)} }
\end{equation}
where $P_{\varphi^{(V)}} \equiv \left< \varphi^{(V)} \right| P \left| \varphi^{(V)} \right>$. At $is = 2/p-1/2$ we have $e^{-2\pi s} = - e^{4\pi i/p}$, so \eqref{eq:rhs} gives:
\begin{equation}
- \partial_{V} D_p(\psi|_{V}) = \frac{p}{p-2} \sin(2\pi/p)\frac{P_{\varphi^{(V)} }}{\| \varphi^{(V)} \|^2} = 2\pi \Theta_p(\psi|_V)
\end{equation}
using \eqref{eq:Thetap}. Comparing to Lemma~\ref{lem:log-convexity} we see the inequalities are saturated as $a \rightarrow 0$, proving the claimed limit $a \rightarrow 0$. Left endpoint convexity follows immediately. 
\end{proof}

Combining Lipschitz continuity of $D_p(\psi |_V)$ with left endpoint convexity gives convexity:
\begin{lemma}
\label{lem:convexity}
Consider a function $f(V)$ defined for $V \geq 0$ which is Lipschitz continuous and satisfies ``left endpoint convexity'' everywhere:
\begin{equation}
(V_2-V_1) \partial_+ f(V_1) \leq f(V_2) - f(V_1)  \qquad \forall V_2 \geq  V_1 \geq 0
\end{equation}
Then $f(V)$ is a convex function.
\end{lemma}
\begin{proof}
Consider $V_\lambda = \lambda V_a + (1- \lambda) V_b$ where $V_{b} > V_a \geq 0$. Define $g(V) = f(V) + \alpha V+ \beta$ where
$\alpha, \beta$ are chosen such that $g(V_a) = g(V_b) = 0$. Linear functions saturate left endpoint convexity so we get the same
inequality for $g(V)$ and all $V \geq 0$. We need to show that $g(V) \leq 0$ for $V_a \leq V \leq V_b$ so that $g(V_\lambda) \leq \lambda g(V_a) + (1-\lambda) g(V_b)$
from which we can clearly replace $g \rightarrow f$ giving convexity. 

Suppose not. Then by continuity there is an open interval $(V_\ell,V_r) \subset (V_a, V_b)$ with $g(V) > 0$ for $V \in (V_\ell,V_r)$ and
with $g(V_\ell) = g(V_r) =0$. Applying the inequality with $V_2 = V_r$ and $V_1 = V \in (V_\ell,V_r)$ and using Lipschitz continuity we can integrate:
\begin{equation}
0 = g(V_r) - g(V_\ell) \leq - \int_{V_\ell}^{V_r} d V \frac{ g(V)}{V_r - V} < 0
\end{equation}
a contradiction. 
\end{proof}

\begin{proof}[Proof of Theorems~\ref{thm:main1} and \ref{thm:main2}] 
We first take $2 < p < \infty$; the cases $p=\infty$ and $p=2$ are treated separately below.
Combining Lemmas~\ref{lem:log-convexity}, \ref{lem:finalderiv} and \ref{lem:convexity} we have most of the statements in the theorems.
We have convexity for $[V_1, \infty)$ for $V_1 > V_0$. But we can use right continuity of $V \rightarrow D_p(V)$
at $V_0$ (apply lower semicontinuity of $D_p$ in the state with the strongly continuous $U(a)\psi$) to shift this back to $[V_0,\infty)$. 
Since $D_p(\psi|_V)$ is convex and differentiable for $V > V_0$, its derivative is continuous there, so $V \rightarrow \Theta_p(\psi|_V) = - \partial_V D_p(\psi|_V)/(2\pi)$ is continuous and non-increasing for $V > V_0$. We also get monotonicity of $\Theta_p(\psi|_V)$ on $[V_0,\infty)$: by Lemma~\ref{lem:log-convexity} and convexity, $\Theta_p(\psi|_{V_0}) \geq (D_p(\psi|_{V_0}) - D_p(\psi|_{V_0+a}))/(2\pi a) \geq - \partial_V D_p(\psi|_{V})|_{V=V_0+a}/(2\pi) = \Theta_p(\psi|_{V_0+a})$ for $a>0$. If we allow $\Theta_p(\psi|_V)$ to be valued in the extended real line, lower semicontinuity of $\Theta_p(\psi|_V)$ from the right (see Lemma~\ref{lem:Thetacont}) with monotonicity gives continuity to $V=V_0$. 

Local Lipschitz continuity gives the integral relation \eqref{mainintegral} for $V_0 < V_1 < V_2$, and this extends to $V_1 = V_0$ by right continuity of $D_p$ at $V_0$ and monotone convergence of the integral. 
The limit in (1) of Theorem~\ref{thm:main1} also holds at $V = V_0$: by Lemma~\ref{lem:Thetacont}, monotonicity and convexity, $\Theta_p(\psi|_{V_0}) \leq \lim_{b \rightarrow 0^+} \Theta_p(\psi|_{V_0+b}) = - \partial^+_V D_p(\psi|_{V})|_{V=V_0}/(2\pi) = \lim_{a \rightarrow 0^+} (D_p(\psi|_{V_0}) - D_p(\psi|_{V_0+a}))/(2\pi a) \leq \lim_{a \rightarrow 0^+} \Theta_p(\psi|_{V_0};a) \leq \Theta_p(\psi|_{V_0})$, using the bounds of Lemma~\ref{lem:log-convexity}. 

For $p=2$, statements (1)-(3) of Theorem~\ref{thm:main2} derive from Theorem 1 of \cite{Faulkner:QNECwithoutAssumptions}.

Now consider statement (4). Continuity in $p'$ of $D_{p'}(\psi|_V)$ is (3) of Lemma~\ref{lem:Dpbasic}. We need continuity in $p'$ for the non-conditional quantum expansion. Consider the relation to the relative entropy from statements (2) and (3) and pick some $2 \leq p_0 \leq p$ and $V \geq V_0$
then:
\begin{equation}
\liminf_{p' \rightarrow p_0}\Theta_{p'}(\psi|_{V}) \geq \sup_{a > 0} \frac{D_{p_0}(\psi|_{V}) - D_{p_0}(\psi|_{V+a})}{2\pi a}
=  \Theta_{p_0}(\psi|_{V})
\end{equation}
implying lower semicontinuity. For $V > V_0$ 
\begin{equation}
\limsup_{p' \rightarrow p_0}\Theta_{p'}(\psi|_{V}) \leq \inf_{V_0 - V< \epsilon < 0} \frac{D_{p_0}(\psi|_{V}) - D_{p_0}(\psi|_{V+\epsilon})}{2\pi \epsilon}
=  \Theta_{p_0}(\psi|_{V})
\end{equation}
implying continuity (the last equality uses the differentiability of $D_{p_0}(\psi|_V)$ at $V > V_0$ from (1)). The same argument applies for $p_0 = \infty$ (when $p=\infty$), where $D_\infty(\psi|_V)$ is constant and $\Theta_\infty = 0$ (Remark~\ref{rem:pinfty}).

For $p = \infty$ all remaining statements are trivial by Remark~\ref{rem:pinfty}: $D_\infty(\psi|_V)$ is constant for $V \geq V_0$ and $\Theta_\infty(\psi|_V) = \Theta_\infty(\psi|_V;a) = 0$.

\end{proof}

\section{Discussion}

In this paper we have completed a general proof of the R\'{e}nyi QNEC - which states that R\'{e}nyi relative entropy is convex along HSMI algebras. 
We introduced several new quantum information quantities to study this problem and worked out various properties of these quantities. 
In particular we focused on properly infinite subalgebras with jointly cyclic and separating vectors, where standard conditional entropies may not be well defined. While there are some known results in quantum information such as approximate Petz sufficiency \cite{junge2018universal} that extend to such subalgebras \cite{Faulkner:2020iou,Junge:2020bwo,Faulkner:2022ada}, we have identified a rich class of new quantities that are worthy of further study.

We finish with some open questions.  The most pressing is to understand the relation between
these new conditional entropies and semiclassical gravity. There are many hints at connections - via their relevance to the QNEC \cite{Bousso:2015mna,Bousso:2015wca}, the limit $p \rightarrow 2$
to the conditional dynamical entropy \cite{Hollands:2024dynamical} and the noted importance of R\'{e}nyi conditional entropies in one-shot holography \cite{Akers:2020pmf,Akers:2023oneshot}. These conditional entropies
are usually defined with respect to inclusions that have normal conditional expectations (such as a normalized partial trace) \cite{Junge:2010mixednorm,Gao:2019subalgebras,Bergh:2026skc,Faulkner:2020hzi} or
normal semifinite faithful (nsf) operator-valued weights \cite{Connes:1980spatial,Cao:2025els}. In certain limits this is probably not the correct
model of an inclusion in semiclassical gravity when QFT dynamics is important, and in particular while tracking quantum information
along null hypersurfaces such as black hole horizons. 
 In this case it is not known whether such normal conditional expectations or nsf operator-valued weights
generically exist, and the physical interpretation for the inclusions considered here remains unsettled. See \cite{Faulkner:2024gst} for further discussion.

While the diagonal R\'{e}nyi QNEC is mathematically true for $p \geq 2$, its physical significance is less clear.
The fact that there is also no statement for $1 \leq p <2$ or the off-diagonal QNEC is rather troubling. It is possible that
the more fundamental objects to study are the weighted conditional entropies. So perhaps monotonicity of the generalized expansions
$\Theta_p(\psi|_V;a)$ with $a$ is the more fundamental statement. 
In upcoming work \cite{Faulkner:WeightedConditionalLp} we extend these norms to $1 \leq p < 2$ where we find various forms of monotonicity and convexity that do extend to this case. 
Using the same logic one might
 guess that the correct R\'{e}nyi generalization of strong subadditivity is monotonicity of the \emph{weighted} conditional entropies $H_2$ (using the analog of SSA written as monotonicity of vanilla conditional entropies); however this seems to fail even at $p=2$ so more thought is required for this.

It is also important to study other families of conditional norms generalizing \cite{Junge:2010mixednorm,Gao:2019subalgebras}. This is a rather involved task because the space of such
norms turns out to be quite large - in particular we now also have a choice of contraction $T_z(\mathcal{A}' \subset \mathcal{B}')$
in generalizing these norms. Physical guidance will be important to sort out the physically interesting ones. For example it would be nice to identify the weighted version of $\min,\max$ conditional entropies \cite{Renner:2005qbj,Koenig:2009avh,Berta:2015nkz}. Another interesting target is the $\alpha$-$z$ QNEC, discussed earlier in \cite{Moosa:2020jwt} and recently studied through escort-state representations in \cite{Kibe:2026dtz,Kibe:2026idb}. More general weighted conditional and amalgamated $L_p$ spaces may provide the appropriate framework for proving this conjecture just as the weighted conditional norms introduced here underpin our proof in the sandwiched case.

It is also an open question to generalize these conditional norms beyond vacuum states to vacuum weights, 
as was recently done for the non-R\'{e}nyi QNEC \cite{Faulkner:QNECwithoutAssumptions}.

\section{Acknowledgements}

We are grateful for discussions with Antony Speranza and Marius Junge. This work is supported by the DOE
award number DE-SC0015655. 

LLMs (Claude Opus 5.5 and GPT-6 Astra) were used at various stages of this work. We acknowledge their help with minor editing, checking and streamlining some proofs and clarifying some edge cases. The main proof ideas were found before the AIpocalypse without the help of A.I. The author takes full responsibility for the work herein. 

\appendix

\section{Proof of the technical lemma}
\label{app:proofs}

\begin{proof}[Proof of Lemma~\ref{lem:technical}]
Note that:
\begin{equation}
 \liminf_{\epsilon \rightarrow 0} \sup_{u' \in \Sigma}  \| \Delta_{\mathcal{B}}^{-\epsilon} \Delta_{\mathcal{A}}^{\epsilon} u' \psi_\epsilon \| 
 \geq  \lim_{\epsilon \rightarrow 0} \left(\| \Delta_{\mathcal{B}}^{-\epsilon} \Delta_{\mathcal{A}}^{\epsilon} x' \psi \| - \| \psi - \psi_\epsilon\| \right)=  \| x' \psi \| >0
 \end{equation}
 where $\psi \notin {\rm ker}(x')$ implying that for small enough fixed $\epsilon > 0$ both $\sup\ln$'s are bounded (they are clearly bounded from 
 above).

We pick $\epsilon \rightarrow u'_\epsilon$ such that the first ($\psi_\epsilon$-dependent) scaled term in the following display, 
evaluated at $u' = u'_\epsilon$, is within $\epsilon$ of its supremum, so that we can bound the second term using this $u'$:
\begin{align}
& \limsup_{\epsilon \rightarrow 0^+} \left( \sup_{u' \in \Sigma} \frac{1}{\epsilon} \ln \| \Delta_{\mathcal{B}}^{-\epsilon} \Delta_{\mathcal{A}}^{\epsilon} u' \psi_\epsilon \| 
-\sup_{u' \in \Sigma} \frac{1}{\epsilon} \ln \| \Delta_{\mathcal{B}}^{-\epsilon} \Delta_{\mathcal{A}}^{\epsilon} u' \psi \| \right)  \\
& \qquad  \leq 
\limsup_{\epsilon \rightarrow 0^+} \left| \frac{1}{\epsilon} \ln \| \Delta_{\mathcal{B}}^{-\epsilon} \Delta_{\mathcal{A}}^{\epsilon} u_{\epsilon}' \psi_\epsilon \| 
-\frac{1}{\epsilon} \ln \| \Delta_{\mathcal{B}}^{-\epsilon} \Delta_{\mathcal{A}}^{\epsilon} u_\epsilon' \psi \| \right|
\end{align}
We note that the $\ln$ term $0 \geq \liminf_{\epsilon \rightarrow 0} \ln \| \Delta_{\mathcal{B}}^{-\epsilon} \Delta_{\mathcal{A}}^{\epsilon} u_{\epsilon}' \psi_\epsilon \|  > -\infty$
is bounded from the discussion just above and also $0\geq \liminf_{\epsilon \rightarrow 0} \ln \| \Delta_{\mathcal{B}}^{-\epsilon} \Delta_{\mathcal{A}}^{\epsilon} u_{\epsilon}' \psi \|  > -\infty$
is bounded because it is close to the aforementioned $\liminf$. 
It follows that we can estimate the log difference using the polynomial difference:
\begin{align}
&\frac{1}{\epsilon}| \| \Delta_{\mathcal{B}}^{-\epsilon} \Delta_{\mathcal{A}}^{\epsilon} u_{\epsilon}' \psi_\epsilon \|^2 - \| \Delta_{\mathcal{B}}^{-\epsilon} \Delta_{\mathcal{A}}^{\epsilon} u_\epsilon' \psi \|^2 | \\ & \qquad
\leq \frac{1}{\epsilon} \left( 2\| \psi_\epsilon - \psi \|^2 + 2 \| \psi_\epsilon - \psi \| \left( 1 -\left< \psi_\epsilon \right| (u_\epsilon')^\dagger \Delta_{\mathcal{A}}^{\epsilon} \Delta_{\mathcal{B}}^{-2\epsilon} \Delta_{\mathcal{A}}^{\epsilon}   u_\epsilon' \left| \psi_\epsilon \right>  \right)^{1/2} \right) \\
&\qquad \leq \frac{1}{\epsilon} \left( 2\| \psi_\epsilon - \psi \|^2 + 2 \| \psi_\epsilon - \psi \| \left( 1 -e^{-2\epsilon^2} \left< \psi_\epsilon \right|(x')^\dagger \Delta_{\mathcal{A}}^{\epsilon} \Delta_{\mathcal{B}}^{-2\epsilon} \Delta_{\mathcal{A}}^{\epsilon} x'  \left| \psi_\epsilon \right>  \right)^{1/2} \right)
\end{align}
where in the last inequality we used the near-optimality of $u_\epsilon'$. 
Now by convergence of $\psi_\epsilon \rightarrow \psi$ we can show that:
\begin{equation}
\lim_{\epsilon \rightarrow 0} 
\left< \psi_\epsilon \right| (x')^\dagger \Delta_{\mathcal{A}}^{\epsilon} \Delta_{\mathcal{B}}^{-2\epsilon} \Delta_{\mathcal{A}}^{\epsilon} x' \left| \psi_\epsilon \right> 
=  1
\end{equation}
by continuity of $z \rightarrow \left< \psi \right| (x')^\dagger \Delta_{\mathcal{A}}^{z} \Delta_{\mathcal{B}}^{-2z} \Delta_{\mathcal{A}}^{z} x' \left| \psi \right>$
in the strip $0\leq {\rm Re}z\leq 1/2$ and boundedness of $\Delta_{\mathcal{B}}^{-z} \Delta_{\mathcal{A}}^{z}$ in that strip. 
And using the assumptions on the rate of convergence $\psi_\epsilon \rightarrow \psi$ we find that:
\begin{equation}
 \limsup_{\epsilon \rightarrow 0^+} \left( \sup_{u' \in \Sigma} \frac{1}{\epsilon} \ln \| \Delta_{\mathcal{B}}^{-\epsilon} \Delta_{\mathcal{A}}^{\epsilon} u' \psi_\epsilon \| 
-\sup_{u' \in \Sigma} \frac{1}{\epsilon} \ln \| \Delta_{\mathcal{B}}^{-\epsilon} \Delta_{\mathcal{A}}^{\epsilon} u' \psi \| \right) \leq 0
\end{equation}
 Reversing $\psi_\epsilon$ and $\psi$ we can go through a similar set of steps (with a different set of $u_\epsilon'$) to complete the two-sided estimate
\eqref{eq:technical}: $\limsup_{\epsilon \rightarrow 0^+} (\cdot) 
\leq 0 \leq \liminf_{\epsilon \rightarrow 0^+}(\cdot)$.

\end{proof}

\bibliography{entanglement}

\end{document}